\documentclass[11pt]{article}

\usepackage{moreverb,url}

\usepackage{latexsym}
\usepackage{amsmath}
\usepackage{color}
\usepackage{graphicx}
\graphicspath{{figures/}}
\usepackage{caption}
\usepackage{subcaption}
\usepackage{amssymb}
\usepackage{float}
\usepackage{longtable}
\usepackage{amsmath}
\usepackage{algorithm} 
\usepackage{algpseudocode}
\usepackage{algorithmicx}
\usepackage{natbib}

\DeclareMathOperator{\E}{\mathbb{E}}
\DeclareMathOperator{\SE}{\text{SE}}

\DeclareMathOperator*{\logit}{logit}

\newcommand{\ctmle}{\text{c-tmle}}
\newcommand{\CV}{\text{CV}}

\newcommand{\bG}{\bar{G}}

\newcommand{\bQ}{\bar{Q}}

\usepackage{enumitem}
\newcounter{descriptcount}

\newtheorem{assumption}{Assumption}

\newcommand{\indep}{\rotatebox[origin=c]{90}{$\models$}}

\setlist[description]{leftmargin=0.25cm,labelindent=0.25cm}

\newcommand\BibTeX{{\rmfamily B\kern-.05em \textsc{i\kern-.025em b}\kern-.08em
T\kern-.1667em\lower.7ex\hbox{E}\kern-.125emX}}

\begin{document}

\title{On Adaptive Propensity Score Truncation in Causal Inference}

\author{Cheng Ju $^1$, Joshua Schwab $^1$, Mark J. van der Laan $^1$}

\date{%
   $^1$ Division of Biostatistics, University of California, Berkeley\\
}

\maketitle
\begin{abstract}
  
The  positivity assumption, or the experimental treatment assignment (ETA) assumption, is important for identifiability in causal inference.  Even if the positivity assumption holds, practical violations of this assumption may jeopardize the finite sample performance of the causal estimator. One of the consequences of  practical violations of the positivity assumption is extreme values in the estimated propensity score (PS). A common practice to address this issue is truncating the PS estimate when constructing PS-based estimators. In this study, we propose a novel adaptive truncation method, Positivity-C-TMLE, based on the collaborative targeted maximum likelihood estimation (C-TMLE) methodology. We demonstrate the outstanding performance of our novel approach in a variety of simulations by comparing it with other commonly studied estimators. Results show that by adaptively truncating the estimated PS with a more targeted objective function,  the Positivity-C-TMLE estimator achieves the best performance for both point estimation and confidence interval coverage among all estimators considered.
  
\end{abstract}


\section{Introduction}


Propensity score (PS) \citep{rosenbaum1983central,rosenbaum1984reducing}, defined as the conditional probability of treatment assignment given measured baseline covariates,  plays an important role in the  estimation of causal effects in observational studies. PS-based methods have been widely  used and studied in the literature, including stratification/sub-classification based on the PS \citep{d1998tutorial,rosenbaum1984reducing}, matching on the PS \citep{rosenbaum1983central}, and weighting the observations with the PS \citep{horvitz1952generalization,rosenbaum1987model}.

The  positivity assumption, or the experimental treatment assignment (ETA) assumption, is important for the identifiability for estimating ATE. The positivity assumption requires  $0 < \bG_0(W)  < 1$ for $W$ almost everywhere, where $\bG_0(W)$ is the PS (the probability to be assigned in the treatment group conditional on the pre-treatment baseline covariate vector $W$). Intuitively, this assumption guarantees that there exist samples in both treatment and control group for each sub-population, so the information for the corresponding potential outcome is available. However, even if the assumption is valid for the true data generating distribution, the randomness in data generating/sampling might cause  practical violation of the positivity assumption. For example, there might be few or even no observations in certain sub-population that are  exposed to treatment. This usually challenges the estimation of the treatment effect for this sub-population. For example, it causes extreme values in the PS estimate, which jeopardizes the performance of the PS based estimators.

Many approaches have been proposed and studied to address practical positivity violations. \cite{petersen2012diagnosing} systematically reviewed several commonly used practice.
One simple and practical method is truncating extreme values in the PS estimate\citep{potter1993effect,scharfstein1999adjusting,cole2008constructing}.  \citep{bembom2008data} proposed an algorithm that selects the truncation level for the inverse propensity score weighted (IPW) estimator by minimizing its estimated mean squared error (MSE).\citep{lee2011weight} further studied the sensitivity of a particular PS weighting estimator of ATE, with the PS estimated by four machine learning algorithms, and truncated at different cutpoints.  Based on \citep{bembom2008data}, \citep{xiao2013comparison} proposed and compared several adaptive truncation methods for marginal structural Cox models. 
Exclusion of problematic  $W$s which result in practical positivity violations (restricting the adjustment set \citep{petersen2012diagnosing}) is another commonly used approach \citep{bembom2008data,petersen2012diagnosing}. While  removing such covariates might increase the bias of the causal estimator from the confounding, it usually substantively reduces the variance.
Sample trimming (restricting the sample \citep{petersen2012diagnosing}), which discard classes of subjects with limited variability in the observed treatment assignment, is another well-studied approach and has been widely used, especially in the econometrics and social science literature \citep{dehejia1999causal,heckman1997matching,lalonde1986evaluating,crump2006moving}.

In this study, we focus on the truncation method to address the practical positivity violation. In practice, the PS score is truncated either by a fixed range (e.g. with absolute value restricted in [0.025, 0.975]), or by a fixed percentile (e.g. with value restricted in [0.1, 0.9] percentile): \citep{kang2016practice} studied the impact of arbitrary  cutoffs of the PS at fixed value for multiple estimators, and  \citep{cole2008constructing,lee2011weight} investigated the bias-variance trade-off with different truncation percentiles for propensity score weighting estimators. However, it is reasonable to believe that such fixed truncation strategy may not be not efficient. As the optimal cutoff depends on the choice of the PS estimator, the choice of the causal estimator, and the observed data, it impossible to know the optimal cutpoint a priori. It is reasonable to believe data-adaptive truncation methods would improve the finite sample performance of the causal estimator. We extend the collaborative targeted maximum likelihood estimation (C-TMLE)  methodology to data-adaptive PS truncation. Developed based on targeted maximum likelihood estimation (TMLE) \citep{van2006targeted}, C-TMLE inherits all the attractive properties of TMLE (e.g. doubly robustness, plug-in estimator) \citep{van2010collaborative}. TMLE has been widely studied and applied in a wide range of topics, including causal inference and genomics \citep{gruber2010application},   survival analysis \citep{stitelman2010collaborative}, and  safety analysis \citep{lendle2013targeted}. \citep{ju2016scalable} proposed scalable C-TMLE by replacing the greedy search in \citep{gruber2010application} with a user-supplied ordering, and applied this to  high-dimensional electronic healthcare data.  \citep{porter2011relative} shows C-TMLE is more robust than TMLE. Recently, \citep{van2017ctmle} developed C-TMLE algorithms for continues tuning parameter, with the general theorem of the asymptotic normality of  the resulting C-TMLE estimators.  Based on this work, \citep{van2017ctmle,ju2017collaborative} further proposed LASSO-C-TMLE, where the PS is estimated by LASSO controlled by C-TMLE, and \citep{ju2017collaborative} demonstrated its performance on high-dimensional electronic health dataset. We simply consider the truncation quantile $\gamma$ as a tuning parameter, and extend the C-TMLE algorithm to select the optimal $\gamma$ for the estimation of the causal parameter.

\section{Brief review of the framework for causal effect estimation}

For simplicity, we model the data generating distribution with a non-parametric structural equation model (NPSEM). Consider each observation, $O_i=(Y_i, A_i, W_i)$, is independently generated from the following data generating system:
\begin{equation*}
  \left\{%
  \begin{array}{l}
    W=f_W(U_W),\\
    A=f_A(W, U_A),\\
    Y=f_Y(A, W, U_Y),
  \end{array},\right.
\end{equation*}
where $f_{W}$,  $f_{A}$   and  $f_{Y}$   are  deterministic   functions  and
$U_W, U_A,  U_Y$ are background  (exogenous) variables. Each observation is drawn from such data generating distribution: first generate $(U_W, U_A, U_Y)$, then generate $W$ based on $U_W$, then determine the treatment assignment $A$ based on $(W,U_A)$. Finally generate outcome $Y$ based on $(A, W, U_Y)$. Consider $A$ is a binary indicator for treatment. Then the potential outcome $(Y_1, Y_0)$ could be obtained by intervening on the treatment $A$ in $f_Y$ with 1 or 0:
\begin{equation*}
  \begin{aligned}
    \label{eq:conter}
    Y_{i0} &= f_Y(A_i = 0, W_i, U_{iy}),\\
    Y_{i1} &= f_Y(A_i = 1, W_i, U_{iy})
    \end{aligned}
\end{equation*}
which implies the consistent assumption:

\begin{assumption}[Consistent Assumption]
  \label{assum:consistent}
  \begin{equation*}
    Y_i = Y_{i,A_i} = Y_{i0}(1-A_i) + Y_{i1}A_i.
  \end{equation*}
\end{assumption}

We consider the target parameter of the average treatment effect (ATE):
\begin{equation*}
  \Psi_0 = \E(Y_1) - \E(Y_0).
\end{equation*}
which could be interpreted as the difference between the expectations of the outcome if all the units received treatment, $\E(Y_1)$, versus if all the units did not receive the treatment, $\E(Y_0)$. We further assume background variables are independent $U_W \indep U_A \indep U_y$, which is a sufficient condition for the conditional randomization assumption:

\begin{assumption}[Conditional Randomization]
  \label{assum:cr}
  \begin{equation*}
    (Y_0, Y_1) \indep A | W.
  \end{equation*}
\end{assumption}

We also need the positivity assumption, or the experimental treatment assignment (ETA) assumption:
\begin{assumption}[The Positivity Assumption]
  \label{assum:pos}
  \begin{equation*}
    0 < \bG_0(W) < 1
  \end{equation*}
  almost everywhere.
\end{assumption}

This assumption means that for each subject in the target population, the probability being assigned to the treatment/control group should be positive. We will discuss assumption \ref{assum:pos} in more detail in subsection \ref{subsec:pos}.

\subsection{The importance for the Positivity Assumption}
\label{subsec:pos}

  The positivity assumption \ref{assum:pos} requires that the probability of assignment to both the treatment group and control group is positive, for any units in the target population. Intuitively, if all the units in a certain sub-population were only assigned to the treatment (control) group, we would never get the information of the  potential outcome corresponding to the control (treatment) group for this sub-population. This leads to the non-identifiability of the ATE of the whole population. \citep{petersen2012diagnosing} studied and discussed the estimator-specific behavior of several widely used estimators when the positivity assumption is violated.

  Even if the positivity assumption holds in the (unknown) true data generating distribution, it is still possible that there are practical violations (or random violations \citep{westreich2010invited}) of the positivity assumption, due to randomness in the data generation.
  For example, consider a case where the probability that subjects in a subgroup receive the treatment is extremely low. Then only very few, or even none of such subjects in a given study sample are observed to receive the treatment, which makes it challenging to make inference for this subgroup \citep{cole2008constructing,xiao2013comparison}.
  \citep{westreich2010invited} illustrated the practical positivity violation by a small observational study of daily aspirin intake for prevention of myocardial infarction, where no one aged 31 to 35 years was exposed by chance. In this case, the information of the potential outcome $Y_1$ for such subpopulation is totally missing.
 
    Practical violations of the positivity assumption can result in poor finite sample performance as it can result in highly influential observations. Consider the case where there was only 1 unit with $W = w$ and low PS of treatment. Then this single individual is now providing all of the information about the potential outcome $Y_0$ in the strata $W = w$. For estimators that rely on the estimation of the conditional response $\E(Y|A,W)$, one of the potential outcomes $Y_a$ is never  observed for some $(a,w)$ and thus may require unreliable extrapolation to regions of $(a,w)$ that are not supported by the data \footnote{If $\bQ_n$ is based on a correctly specified parametric model, this extrapolation will be accurate. However, in general we don’t have correctly specified parametric models}.
    For weighting based estimators, this individual usually gets a large weight, which leads to the high variance of the resulting causal estimator.
    In this study, we propose a novel algorithm that provides a stable estimation of the causal parameter when there exists extreme values in the estimated PS due to the practical violation of the positivity assumption.

\subsection{Notations}

We first use $Q(W)$  to denote the marginal distribution, $P(W)$, of $W$; $\bG(W)$ to denote the conditional expectation of $A$ given $W$, $\E(A | W)$, and  $\bQ(A,W)$ to denote conditional expectation of $Y$ given $(A,W)$, $\E(Y | A, W)$. We use $Q_0$, $\bG_0$, and $\bQ_0$ for the corresponding part in the true data generating distribution $P_0$ of $O_i$, and use $Q_n$, $\bG_n$ and $\bQ_n$ to denote the corresponding estimate trained on the whole observed data.

For simplicity, we introduce two loss functions. The first one, $L_1$, is defined for the conditional outcome $\bQ_0$. One example of the loss function for the estimate $\bQ$ with outcome $Y \in [0,1]$ is:
\begin{equation*}
  L_1(\bQ)(O_i) = - \left( Y_i \log(\bQ(W_i))+ (1-Y_i) \log(1-\bQ(W_i)) \right)
\end{equation*}
The second one, $L_2$, is defined for the propensity score $\bG_0$. One example of the loss function for the estimate $\bG$ with binary treatment indicator $A$ is:
\begin{equation*}
  L_2(\bG)(O_i) = - \left( A_i \log(\bG(W_i)) + (1-A_i) \log(1-\bG(W_i)) \right)
\end{equation*}

In addition, we use $\hat{\bG}_{\gamma}(P_{n}^0)$ to denote the resulting PS estimate by fitting estimator $\hat{\bG}$ (e.g. main term logistic regression) of $\bG_0$ on the training data with the given empirical distribution $P_n^0$, and truncated at $\gamma$ percentile. Notice we have $\bG_{n,\gamma} = \hat{\bG}_{\gamma}(P_{n})$, where $P_n$ is the empirical distribution of all the observed units. We directly use empirical distribution $Q_n$ to estimate $Q_0$.

\section{Data-adaptive Truncation}

A consequence of PS truncation is the introduction of bias in the estimated PS, which in turn causes bias in PS-based causal estimators \citep{xiao2013comparison}. Thus, PS truncation requires a bias-variance trade-off: too much truncation can make estimators more stable but also introduce more bias \citep{cole2008constructing,bembom2008data}.
\citep{cole2008constructing} studied the bias-variance trade-off of the PS truncation by progressively truncating the PS weights at different quantiles. However, the optimal truncation varies for different datasets, and is usually unknown. Thus, it is important to define an empirical metric to select the cutpoints for truncation in a data-adaptive manner. Ideally, the optimal cutpoints should be selected minimizing the loss function (e.g. MSE) of the resulting causal estimator. However, the true MSE is not accessible in practice. \citep{bembom2008data} proposed a closed-form estimate for the expected MSE of a truncated IPTW estimator. However, it is difficult to generalize this closed-form MSE estimator to TMLE.

In this study, we propose a data-adaptive method to select the quantile for truncating the PS estimate specially designed for TMLE. In subsection \ref{subsec:cv}, we first describe the commonly used CV-selector for cutpoint selection. In subsection \ref{subsec:ctmle}, we discuss the drawbacks of a model-free CV-selector, and present the Positivity-C-TMLE algorithm for cutpoint selection.

For simplicity, we only consider the case where the practical violation of positivity is one-sided. In other words, if we use inverse propensity score weighted (IPW) estimator, almost all the extreme weights are from the units in the control group where the estimated PSs are close to 1. In this case, we only consider the one-side truncation, which could be defined as:

\begin{equation*}
\bG_{n,\gamma}(W_i) = \min(\bG_n(W_i), q_{\gamma}(\bG_n))
  \end{equation*}
where $q_{\gamma}(\bG_n)$ is the $\gamma$ quantile for the empirical distribution of $\bG_n$.

Notice the framework for one-side truncation could be easily extended to two-side truncation, by adaptively selecting two truncation points. 

\subsection{Data-adaptive Truncation with Cross-validation for $\bG_0$}

\label{subsec:cv}

One of the most straightforward methods to select the cutpoint is  cross-validation. Consider $V$-fold cross validation. Let $B_n\in  \{0,1\}^n$ be a  cross-validation scheme. Define  $B_{n}$ to  be a  $V$-fold cross-validation  scheme,  {\em i.e.},  a random  vector with  $V$ different potential  values,  each  with probability  $1/V$. For each potential value of the random vector $B_{n}$, the sum of all the coordinates is around $n/V$, and across all the potential values for $B_{n}$, the sum of a given coordinate is $1$. Let $P_{n,B_n}^0$  be  the  empirical  probability distribution  of  the  training subsample $\{O_i  : B_n(i)=0,  1  \leq  i \leq n\}$  and $P_{n,B_n}^1$  be the empirical    probability   distribution    of    the   validation    subsample $\{O_{i}: B_n(i)=1, 1 \leq i \leq n\}$.  The cross-validation selector of $\gamma$ is then defined as

\begin{equation*}
  \label{eq:cv}
  \gamma_{n,\CV}   \equiv  \mathop{\arg\min}_{\gamma   \in   \Gamma}  E_{B_n}   P_{n,B_n}^1
  L_2(\hat{\bar{G}}_{\gamma}(P_{n,B_n}^0))
\end{equation*}
where $\Gamma$ is the set of potential cutpoints $\gamma$. $L_2$ can be any binary loss function, and in this study we used a commonly used one, the negative log-likelihood loss function:
\begin{equation*}
  L_2(\bG)(O_i) = - \left( A_i \log(\bG(W_i)) + (1-A_i) \log(1-\bG(W_i)) \right).
\end{equation*}

\subsection{Data-adaptive Truncation by the Stability of $\Psi_n$}
\label{subsec:mv}

The CV-selector for $\bG_0$ has the following drawbacks:

\begin{itemize}
\item The objective function of CV  merely focuses on the predictive performance of $\bG$. In other words, it does not apply any knowledge of the target parameter. 

\item In addition, such CV procedure is ``model-free''. It selects the cutpoint independently (without regard to the causal parameter/estimator), and then plugs the resulting estimate of $\bG_0$ into the estimator of the causal parameter. It is reasonable to believe different estimators of different causal parameters might have different optimal cutpoints. For example, the vanilla IPW estimator might need more truncation (lower cutpoint in our case), compared to the stabilized Hajek-type IPW.
\end{itemize}

To overcome this, it is important to consider a better empirical metric on the parameter of interest (e.g. MSE for the causal parameter). However, this is hard to achieve, as the value of the causal parameter is unknown. \citep{bembom2008data} proposed an closed-form estimate of the MSE:
\begin{equation*}
\text{MSE}(\gamma) = \text{Var}(\gamma) + \text{Bias}^2(\gamma)
 \end{equation*}
of the IPW estimator at the truncation level $\gamma$, and use it to select the truncation level. However, this closed-form estimate is hard to extended to more complicated estimators, like TMLE. \citep{xiao2013comparison} extended this work by a repeated two-fold cross-validation approach: the first part of the MSE estimate, $\text{Var}(\gamma)$, is estimated by the variance estimate of the causal estimator. The second part,  $\text{Bias}^2(\gamma) = (\Psi_{n,\gamma} - \Psi_{0})^2$, is estimated by the following procedure:

\begin{enumerate}
\item Randomly split data into two disjoint halves.

\item Compute $\Psi_{n,\gamma}$ on one of the halves with truncation level $\gamma$, and compute $\tilde{\Psi}$ on the other data.

\item Use $\widehat{\text{Bias}}^2(\gamma) = (\Psi_{n,\gamma} - \tilde{\Psi})^2$ to estimate $\text{Bias}^2(\gamma)$.
\end{enumerate}

\citep{xiao2013comparison} suggested repeating the above procedure $k$ times and taking the average of the bias estimates to stabilize the result.

Note the authors called this procedure ``cross-validation'' . To distinguish it from the conventional CV procedure mentioned in the previous subsection, we call it multi-view validation (MV) in our paper.

\subsection{Data-adaptive Truncation with Collaborative Targeted Learning}
\label{subsec:ctmle}

In this subsection, we propose a new algorithm called Positivity-C-TMLE. It is specially designed for the TMLE estimator. We first introduce targeted minimum loss-based estimation, and then discuss this novel algorithm with  details.

\subsubsection{Brief review of Targeted Minimum Loss-based Estimation (TMLE)}

Targeted minimum loss-based estimation (TMLE) is a general methodology to estimate a user-specified parameter of interest \citep{van2006targeted}. TMLE estimator is  double robust, which means it is consistent as long as at least one of $\bG_n$ and $\bQ_n$ is consistent. In addition, TMLE estimator is efficient if both the input estimator $\bG_n$ and $\bQ_n$  are consistent.

In this study, we consider the TMLE for estimation of ATE, with the negative likelihood as the  loss function, and logistic fluctuation. Then the TMLE algorithm can be written as:


\begin{algorithm}[H]
  \caption{TMLE Algorithm for ATE, with negative log-likelihood loss  and logistic fluctuation}
  \label{algo:tmle}
  \begin{algorithmic}[1]
    \Function{TMLE}{$\bar{Q}_n^0$, $\bG_n$, $P_n$}
    \State{Construct clever covariate $H_{\bG_n}(A_i, W_i) = \frac{A_i}{\bG_n(W_i)} - \frac{1 - A_i}{1- \bG_n(W_i)}$}
    \State{Fit a logistic regression: the outcome is $Y_i$, with $\bQ_n^0$ as intercept, and  $H_{\bG_n}(A_i, W_i)$ as the univariate predictor, with coefficient $\epsilon$.}
    \State{Fluctuate the initial estimate: Given the logistic model above with fitted coefficient $\epsilon$ of $H_{\bG_n}$, update the initial estimate by
$$\logit(\bQ_n^*(A,W)) = \logit(\bQ_n^0(A,W)) + \epsilon H_{\bG_n}(A,W)$$}

          \Return{$\bar{Q}_n^{*}$}
          \EndFunction
  \end{algorithmic}
\end{algorithm}
Then the resulting TMLE estimator for the ATE can be written as:

\begin{equation}
  \label{eq:tmle}
\Psi_n^{TMLE} = \frac{1}{n}\sum_{i=1}^{n} (\bar{Q}_n^*(1, W_i) - \bar{Q}_n^*(0, W_i)).
\end{equation}

To construct a good input $\bar{Q}_n^0$ and $\bG_n$ for algorithm \ref{algo:tmle}, we suggest  using Super Learner, a  cross-validation based ensemble learning method. Super Learner could easily combine a set of individual machine learning algorithms, and has demonstrated outstanding performance in a wide range of tasks, including causal inference \citep{pirracchio2015improving,gruber2015ensemble,rose2016machine,ju2016propensity}, spatial prediction \citep{davies2016optimal}, online learning \citep{benkeser2017online}, and image classification \citep{ju2017relative}. We refer the interested reader  to the literature on Super Learner \citep{van2007super,polley2010super}.

In addition to the double robustness and asymptotic efficiency mentioned above, TMLE has following advantages:

\begin{enumerate}
\item Equation \ref{eq:tmle} shows that TMLE is a plug-in estimator, which respects the global constraints of the model by mapping the targeted estimate $P^*$ (defined by $(\bQ_n^*,Q_n)$) of $P_0$ into the target parameter $\Psi$. Note some other estimators (e.g. IPW) may produce estimates out of such constrains. 
\item The loss function defined in TMLE, $L_1$ (the negative log-likelihood loss in algorithm \ref{algo:tmle}), offers a metric to evaluate the goodness-of-fit of $(\bG_n, \bar{Q}_n)$, directly w.r.t. the parameter of interest $\Psi_0$. In this example, the loss is the negative log-likelihood in the logistic regression step of algorithm \ref{algo:tmle}.
  \item \citep{porter2011relative} used simulations to compare TMLE, IPW, and A-IPW.  Simulation results show TMLE is the most robust  to (near) positivity violations among three estimators.
\end{enumerate}

\subsubsection{Positivity-C-TMLE}
\label{subsubsec:ctmle}

In this section, we introduce the Positivity-C-TMLE algorithm, which is based on the general template of C-TMLE \citep{van2010collaborative,ju2016scalable}. The basic philosophy of C-TMLE is to sequentially generate a sequence of $\bQ_{n,k}^*$ indexed by $k$, each corresponding to a $\bG_{n,k}$, where the empirical loss $P_nL_2$ for  $\bG_{n,k}$ is monotonically deceasing. The input of the C-TMLE algorithm is a user-provided initial estimate $\bar{Q}_n^0$ for $\bar{Q}_0 = \E_0(Y | A, W)$  with the empirical distribution $P_n$ of the observed data $O_i, i = 1, \ldots, n$. Algorithm \ref{algo:ctmle} shows the details of C-TMLE algorithm for cutpoint selection. Note the tuning parameter, the cutpoint $\gamma$, could be easily replaced by any one-dimensional tuning parameter (e.g. regularization parameter $\lambda$ for LASSO \citep{ju2017collaborative,van2017ctmle}). For simplicity, we first introduce algorithm \ref{algo:ctmle_construct} for construction of the C-TMLE candidate:

\begin{algorithm}[H]
  \caption{Positivity-C-TMLE Candidate Construction Algorithm}
  \label{algo:ctmle_construct}
  \begin{algorithmic}[1]
    \Function{Generate-Candidates}{$\bar{Q}_n^0$, $P_n$, $\Gamma_k$, $\gamma_{\min} = 0.6, \gamma_{\max} = 1$}\
    \State{Train an estimator $\bG_n$ of $\bG_0$ on $P_n$.}
    \State{Construct a sequence of propensity score model
      $\bG_{n,\gamma}$ indexed by the corresponding cutpoint $\gamma$, where $\gamma$ is the $\gamma$-th empirical quantile of the estimated PS.}
    \State{We further set $\gamma$ within the set
      $\Gamma = [\gamma_{\min}, \gamma_{\max}]$.}
    \State{Initialize $k = 1$, $\gamma_0 = 1$}
    \State{Initialize $\text{Search} = \text{False}$}
    \If{$\Gamma_k$ is not provided}
    \State{Set $\text{Search} = \text{True}$}
    \State{Set $\Gamma_k = []$}
    \EndIf
    \While{$\Gamma$ is not empty}
    \If{\text{Search}}
    \State{Apply targeting
      step for each $\bG_{n,\gamma}$, with $\gamma \in \Gamma$, with initial estimate
      $\bar{Q}_n^k$ and clever covariate $H_{\bG_{n,\gamma}}(A,W)= \frac{1-A}{1
        -\bG_{n,\gamma}(W) } + \frac{A}{\bG_{n,\gamma}(W)}$.}
    \State{Select
      $\gamma_k$ corresponding to the $\bar{Q}_{n,\gamma_k}^*$ with the smallest empirical risk $P_n L_1(\bar{Q}_{n,\gamma_k}^*(A,W))$.}
    \Else
    \State{$\gamma_k = \Gamma_k[k]$}
    \EndIf
    \State{For $\gamma \in [\gamma_{k-1}, \gamma_{k}]$,
      compute the corresponding TMLE using initial estimate $\bar{Q}_n^{k-1}$ and
      propensity score estimate $\bG_{n,\gamma}$. We denote such estimate with
      $\bQ_{n,\gamma}^*$ and record them.}
    \State{Set a new initial estimate $\bar{Q}_n^{k} = \bar{Q}^*_{n, \gamma_k}$.}
    \State{Set $\Gamma =
      (\gamma_k, \gamma_{max}]$.}
      \State{Append $\gamma_k$ to  $\Gamma_k$.}
      \State{Set $k = k + 1$.}
      \EndWhile
      \If{$\text{Search}$}
      
      \Return{$[\bar{Q}_{n, \gamma}^{*}, \gamma \in \Gamma]$}
      \Else

      \Return{[$\bar{Q}_{n,\gamma}^{*}, \gamma \in \Gamma]$, $\Gamma_k$}
      \EndIf
      \EndFunction
  \end{algorithmic}
\end{algorithm}

Algorithm \ref{algo:ctmle_construct} builds a sequence of candidates for the conditional response, $\bQ_n^*$, given the empirical distribution of observed data $P_n$, the set of fluctuation points $\Gamma_k$, and an initial estimate $\bQ_n^0$. 

The empirical loss decreases at each fluctuation point. During the cross-validation step, the set of fluctuation points is given (precomputed by whole training sample), so it would only update $\bQ_n$ at each given fluctuation point. Then C-TMLE uses a targeted CV to select the stopping point:
  
\begin{algorithm}[ht]
  \caption{Positivity-C-TMLE Algorithm}
  \label{algo:ctmle}
  \begin{algorithmic}[1]
    \Function{Positivity-C-TMLE}{$\bar{Q}_n^0$, $P_n$, $\gamma_{\min} = 0.6, \gamma_{\max} = 1, V = 5$}
    \State{Build a sequence of candidates using the whole dataset:
      
      $[\bar{Q}_{n,\gamma}^{*}, \gamma \in \Gamma], \Gamma_k = \text{GENERATE-CANDIDATE}(\bar{Q}_n^0, P_n)$}
    \State{Given the set of the fluctuation points $\Gamma_k$ from the previous step, compute the V-fold CV risk for each candidate:

      Build sequence of candidates $\bQ_{\gamma}^*(P_n^0)$ on training set $P_n^0$ with given $\Gamma_k$ by calling

      $\bQ_{\gamma}^*(P_n^0) = \text{GENERATE-CANDIDATE}(\bar{Q}_n^0, P_n^0, \Gamma_k)$.

       Repeat this for $V$ folds and compute the average validation
       \begin{equation*}
         \label{eq:ctmlecv}
         E_{B_n} P_n^1 L_1(\bQ_{\gamma}^*(P_n^0)).
         \end{equation*}}
        \State{Select the best candidate $\bar{Q}_{n,\gamma_{ctmle}}^*$ among $\bar{Q}_{n,\gamma}^*$, with the smallest cross-validated loss in \eqref{eq:ctmlecv}, and its corresponding initial estimate $\bar{Q}_{n,\gamma_{ctmle}}$.}
      \State{Apply targeting step to $\bar{Q}_{n,\gamma_{\ctmle}}$ from the previous step, with each $g_{n,\gamma}$, $\gamma \in [\gamma_{\min},\gamma_{\ctmle})$, yielding a new sequence of estimate $\bar{Q}_{n,\gamma}^*$.}
        \State{Select $\bar{Q}_n^{*} = \arg\min_{\bar{Q}_{n,\gamma}^{*}}P_n L_1(\bar{Q}_{n,\gamma}^{*}), \gamma \in [\gamma_{\min},\gamma_{\ctmle})$  with the smallest empirical loss from the sequence in last step as the final estimate.}

          \Return{$\bar{Q}_n^{*}$}
          \EndFunction
  \end{algorithmic}
\end{algorithm}
Positivity-C-TMLE algorithm first computes the set of the fluctuation points and a sequence of candidates using the entire dataset, then uses the V-fold CV to compute the CV-loss for each candidate.  It picks up the one with the smallest CV-risk, with corresponding initial estimate $\bar{Q}_{n,\gamma_{\ctmle}}$. Then it fluctuates   $\bar{Q}_{n,\gamma_{\ctmle}}$ with each $\gamma > \gamma_{\ctmle}$, and selects the one $\bar{Q}_n^{*}$ with the smallest empirical loss. The estimator for the causal parameter, ATE, is given by:

\begin{equation*}
  \psi_n^{\ctmle} = \frac{1}{n}\sum_{i=1}^{n}(\bar{Q}_n^{*}(1, W_i) - \bar{Q}_n^{*}(0, W_i))
\end{equation*}

For simplicity, C-TMLE in later sections also refers to the Positivity-C-TMLE described above.

\subsubsection{Remark}

Though it seems complicated at first glance, the Positivity-C-TMLE simply repeatedly:

\begin{itemize}
\item Finds the best $\bG_{n, \gamma}$ to update the current estimate $\bQ_n^k$, with respect to the empirical predictive performance for $\bQ_{0}$.
\item Updates the current estimate $\bQ_n^k$  with selected $\bG_{n, \gamma}$, resulting in candidate TMLE estimator $\bQ_{n, \gamma}^*$. 
\end{itemize}
and finally uses CV to select the best $\bQ_{n,\gamma}^*$ with respect to the cross-validated predictive performance for $\bQ_{0}$.

\section{Experiment}
\label{sec:sim}

In this section, we designed simulation studies to assess the performance (bias, variance, and MSE) of several commonly used estimators. For each estimator, we studied different methods to determine cutpoint. Subsection \ref{subsec:dgd} presents how data was generated for experiments. Subsection \ref{subsec:est} reviews the  estimators used in the experiments. Subsection \ref{subsec:res} shows the results from the simulation, and compares the estimators with different empirical metrics for cutpoint selection. The R package \textit{ctmle} \citep{ju2017package} can be found at \url{https://github.com/jucheng1992/ctmle}.

\subsection{Data Generating Distribution}
\label{subsec:dgd}

We consider the following data generating distribution for $O_i = (Y_i, A_i, W_i)$: $W_i$ is the vector of potential confounders, generated from weakly correlated multivariate normal distribution. The treatment indicator variable $A_i$ is independently generated from a Bernoulli distribution, with:
\begin{equation*}
  P(A_i = 1|W_i) = - (W_{i1} + W_{i2} + \sum_{j=2}^{20}\frac{3}{20}W_{ij} + C).
\end{equation*}
Thus, the PSs would be closer to 1 with a larger value of  intercept $C$.

\begin{figure}[H]
  \centering
  \begin{subfigure}[b]{0.4\textwidth}
    \includegraphics[width=\textwidth]{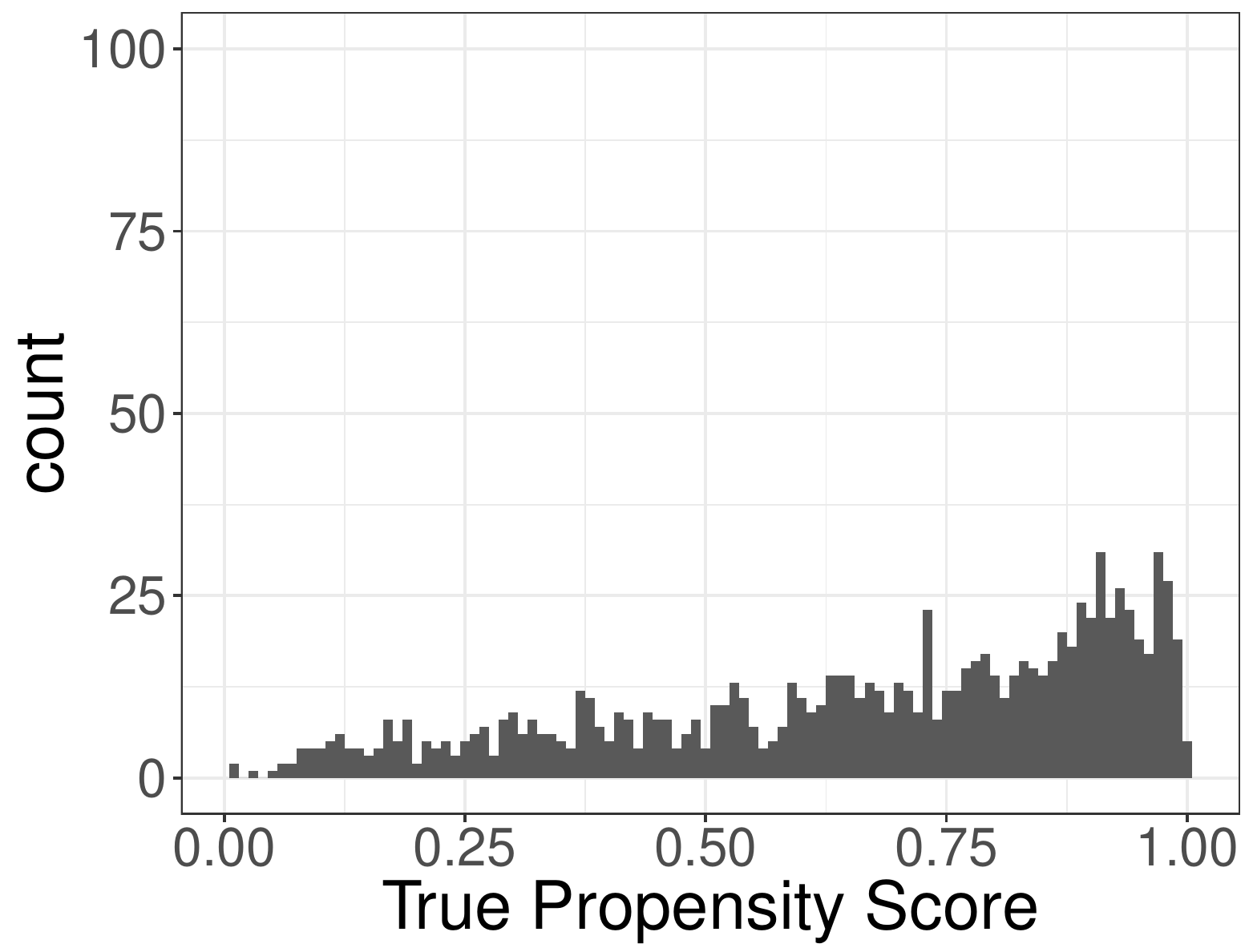}
  \end{subfigure}
  \begin{subfigure}[b]{0.4\textwidth}
    \includegraphics[width=\textwidth]{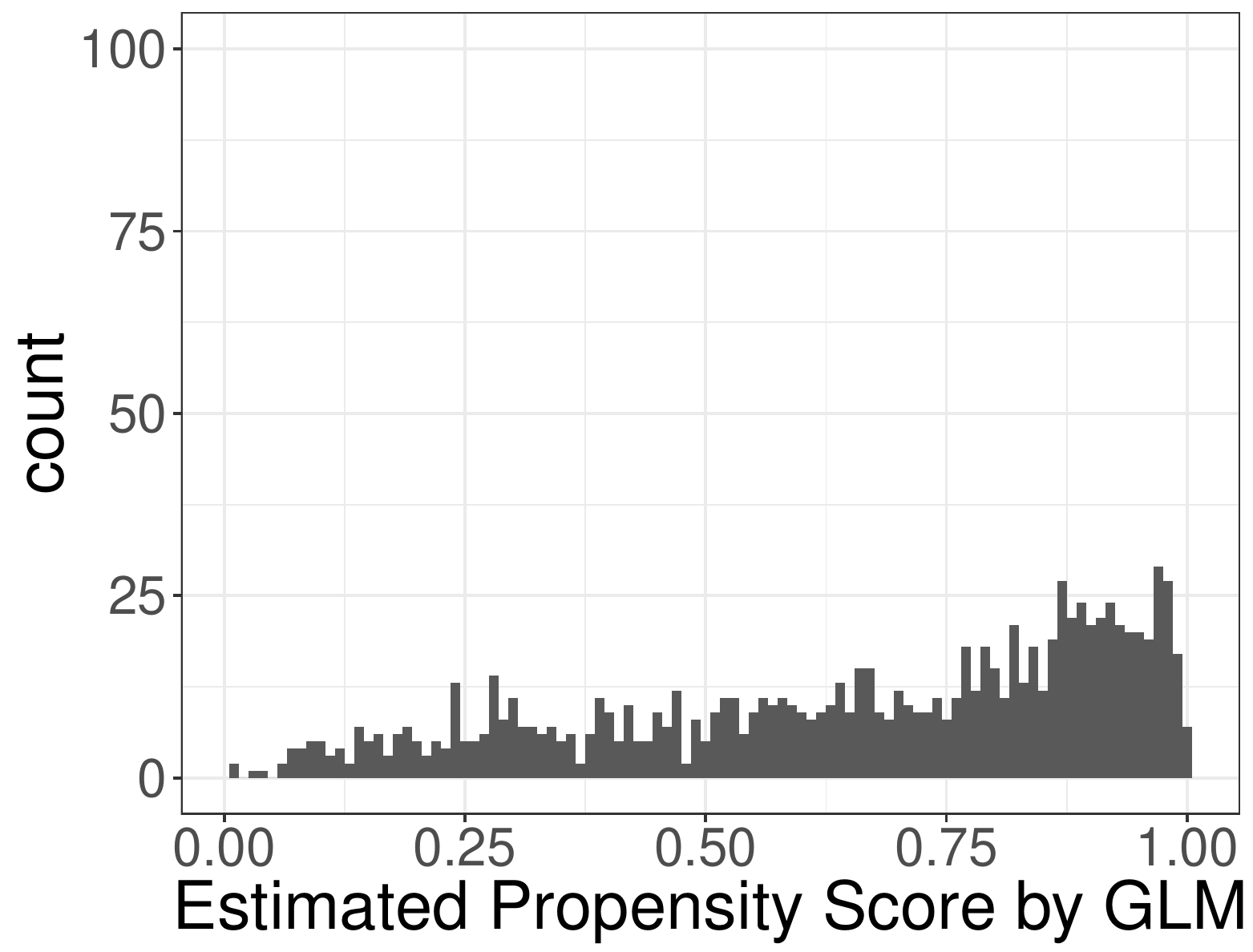}
  \end{subfigure}
  \caption{Histogram for the true PS and estimated PS for C = 1, Sample size N = 1000}\label{fig:g0gnC1}
\end{figure}

\begin{figure}[H]
  \centering
  \begin{subfigure}[b]{0.4\textwidth}
    \includegraphics[width=\textwidth]{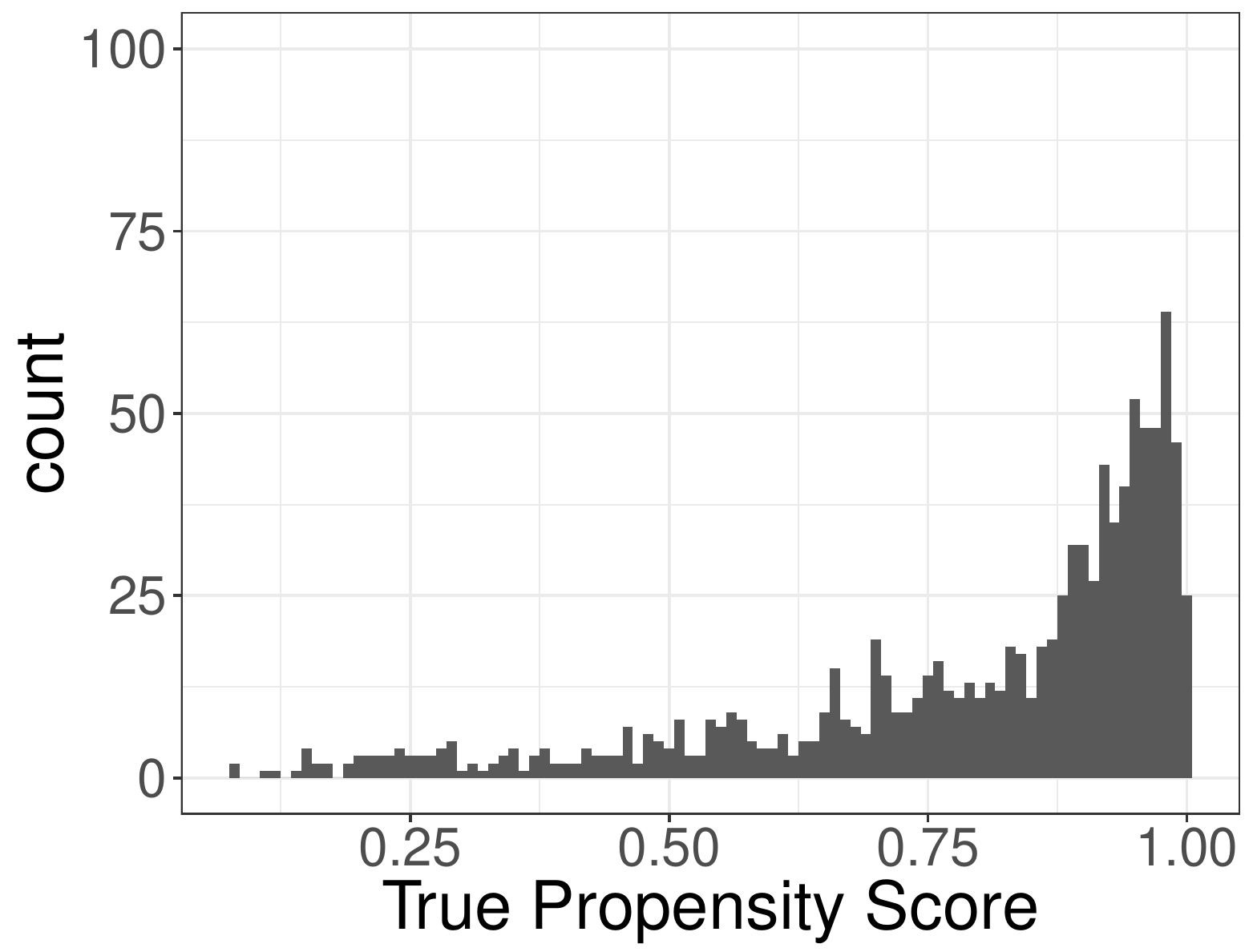}
  \end{subfigure}
  \begin{subfigure}[b]{0.4\textwidth}
    \includegraphics[width=\textwidth]{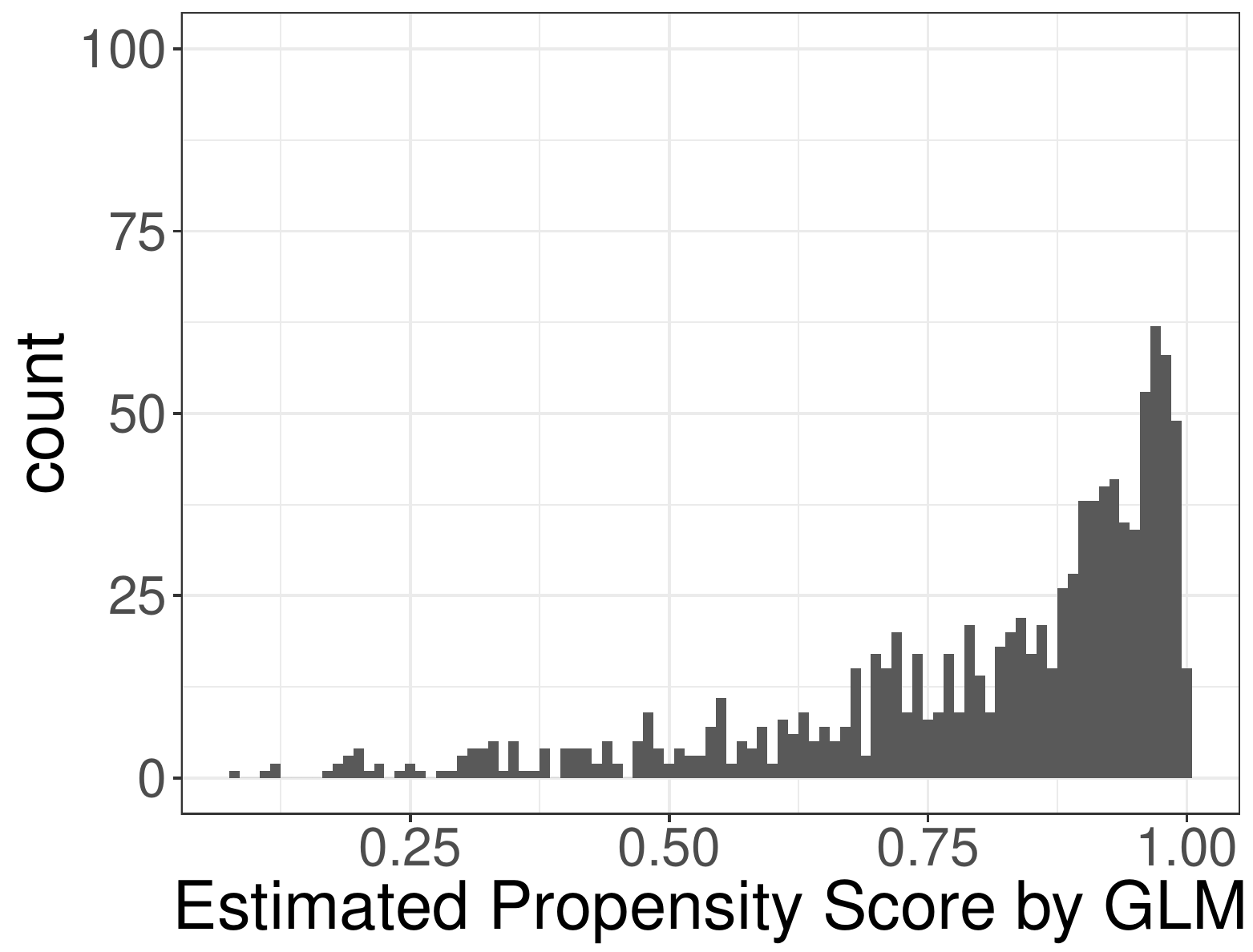}
  \end{subfigure}
  \caption{Histogram for the true PS and estimated PS for C = 2, Sample size N = 1000}\label{fig:g0gnC2}
\end{figure}

Figure \ref{fig:g0gnC1} and \ref{fig:g0gnC2} shows the histogram plots of true propensity score, and estimated propensity score (by logistic regression) for $C = 1, 2$, with sample size $N = 1000$.

The potential outcomes pair $(Y_{i0},Y_{i1})$ is independently generated from a Gaussian distribution, with conditional expectations:

\begin{equation*}
  \E(Y_{i0} \mid W_i) = 2 + 2 (W_{i1} + W_{i2} + W_{i5} + W_{i6} + W_{i8})
\end{equation*}
and
\begin{equation*}
  \E(Y_{i1} \mid W_i) = 4 +  2 (W_{i1} + W_{i2} + W_{i5} + W_{i6} + W_{i8})
\end{equation*}
and the variance are 1 for both $Y_{i0}$ and $Y_{i1}$. In other words, the observed outcome $Y_i$ is from a normal distribution with variance 1 and expectation:
\begin{equation*}
  \E(Y_{i} \mid A_i, W_i) = 2 + 2 (W_{i1} + W_{i2} + W_{i5} + W_{i6} + W_{i8}) + 2 A_i
\end{equation*}

\subsection{Estimators}
\label{subsec:est}

In the simulation, we compared several PS-based estimators. First, we consider the widely used the inverse propensity score (IPW) estimator, or so-called Horvitz-Thompson estimator \citep{horvitz1952generalization}:
\begin{equation*}
  \Psi_n^{IPW} = \frac{1}{n}\sum_{i=1}^{n} \left[\frac{A_iY_i}{\bG_n(W_i)} - \frac{(1-A_i)Y_i}{1- \bG_n(W_i)} \right].
\end{equation*}
IPW is a consistent estimator when $\bG_n$ consistently estimates $\bG_0$. However, due to the inverse weighting, the IPW estimator usually has overly large variance, when there exist some weights $A\bG_n + (1-A)(1-\bG_n)$ close to zero. To stabilize the IPW estimator, the Hajek-type IPW (Hajek-IPW) \citep{hajek1971comment} was proposed as:
\begin{equation*}
  \Psi_n^{Hajek-IPW} = \sum_{i=1}^{n} \left[\frac{A_iY_i/\bG_n(W_i)}{\sum_{i=1}^{n} A_i/\bG_n(W_i)} - \frac{(1-A_i)Y_i/(1- \bG_n(W_i))}{\sum_{i=1}^{n}(1-A_i)/(1- \bG_n(W_i))} \right].
\end{equation*}

Hajek-type IPW is usually more stable compared to the plain IPW estimator. However, this stabilized IPW estimator  will still be highly variable and will have a positively skewed distribution if there are very strong covariate-treatment associations \citep{hernan_brumback_robins,xiao2013comparison,neugebauer2005prefer}.

Both of the above estimators only rely on estimation of the PS and will be inconsistent if the PS is not estimated consistently. We further compared the following double robust estimators. The Augmented-IPW (A-IPW, or DR-IPW) estimator \citep{robins1995semiparametric} can be written as:
\begin{equation*}
  \label{aipw}
  \Psi_n^{DR-IPW} = \frac{1}{n}\sum_{i=1}^{n} H_{\bG_n}(A_i,W_i)\left[Y_i - \bar{Q}_n(A_i,W_i)\right]
  + \bar{Q}_n(1,W_i) - \bar{Q}_n(0,W_i)
\end{equation*}
where
\begin{equation*}
  H_{\bG_n}(A,W) = \frac{A}{\bG_n(W)} - \frac{1-A}{1 - \bG_n(W)} .
\end{equation*}
In this study, we also consider the vanilla TMLE estimator:
\begin{equation*}
  \label{eq:tmle1}
  \Psi_n^{TMLE} = \frac{1}{n}\sum_{i=1}^{n} (\bar{Q}_n^*(1, W_i) - \bar{Q}_n^*(0, W_i)).
\end{equation*}

We consider the following estimators to estimate the causal parameter:

\begin{itemize}

\item Estimators with fixed truncation level: for all the estimators described above, we provided them with estimated PS truncated at different fixed percentile, from $\gamma=60\%$ quantile, to $\gamma=100\%$ quantile (no truncation), with step size $1\%$.

\item Estimators with the truncation level selected by CV: for all the estimators described above, the truncation level for the PS estimate is selected by CV (see details in subsection~\ref{subsec:cv}).

\item TMLE estimator with truncation level selected by MV, or for short, MV-TMLE estimator (see details in subsection~\ref{subsec:mv} ).

\item Positivity-C-TMLE estimator.

\end{itemize}

For  A-IPW, TMLE and C-TMLE estimators which rely on the estimation of $\bQ_0$, we used the estimate $\bQ_n^0$ from a main terms linear regression, with observed outcome, $Y$, as dependent variable, and treatment, $A$, along with baseline covariates $W_3, \ldots, W_{10}$ as predictor. In other word, the confounding in the initial estimate is partially controlled. For the estimation of $\bG_0$ for all PS-based estimators, we used a main terms logistic regression with all the covariates as predictors. In other words, the PS is estimated consistently and efficiently. Thus we guarantee the model is correctly specified, and  the failure of the estimators in the simulations are from the practical violation of the positivity assumption instead of model misspecification.

For each of the following simulation settings, we generated the data from each corresponding data generating system $200$ times, and report the average bias, standard error, and mean squared error of all the estimators.

\subsection{Results}
\label{subsec:res}

We use solid curves with different color to denote the estimators with different fixed quantiles as cutpoint for truncation. For all estimators with data-adaptive truncation, we use horizontal lines to present the performance.

\subsubsection{Mean Squared Error}

\begin{figure}[H]
  \centering
  \begin{subfigure}[b]{0.4\textwidth}
    \includegraphics[width=\textwidth]{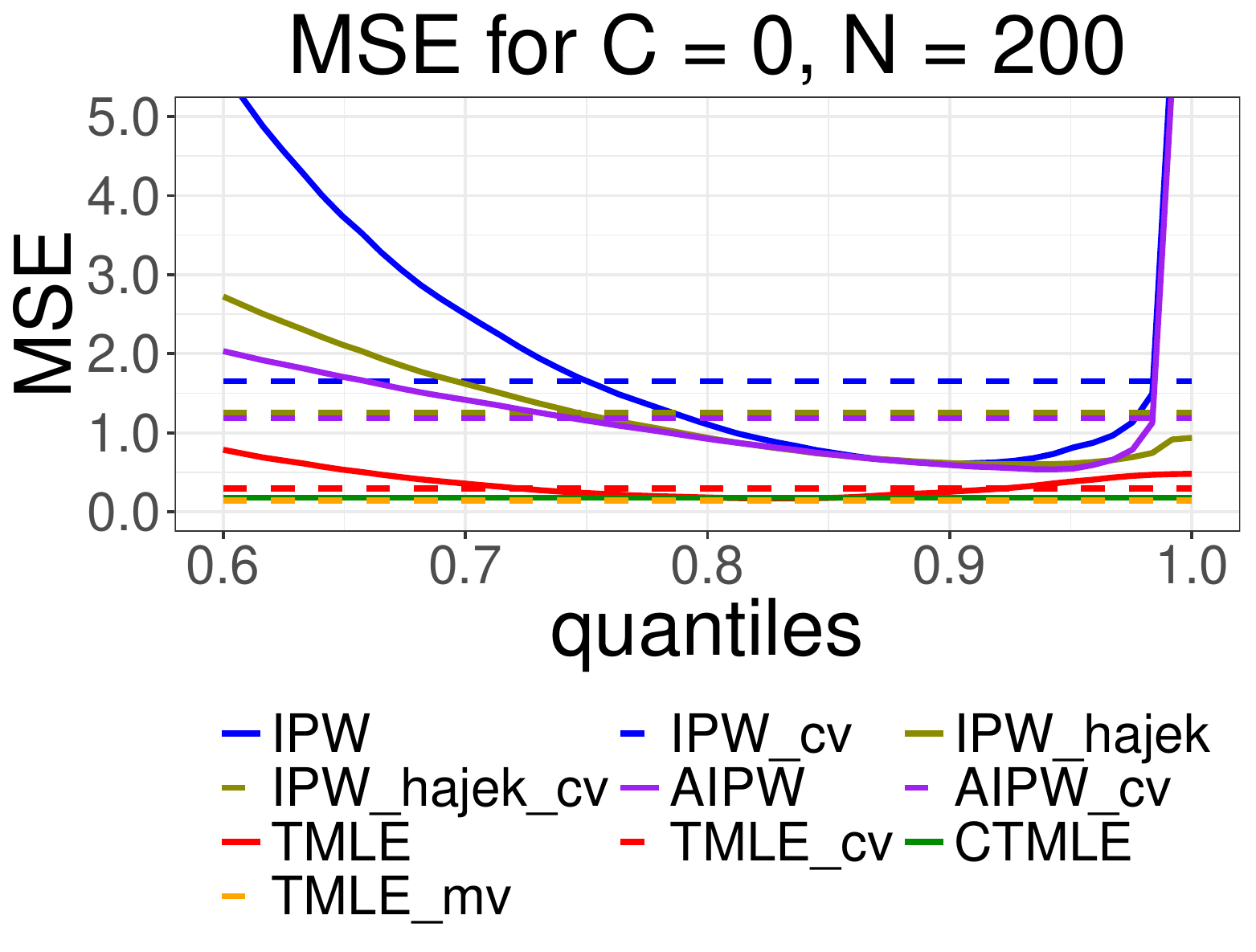}
  \end{subfigure} \hspace{5mm}
  \begin{subfigure}[b]{0.4\textwidth}
    \includegraphics[width=\textwidth]{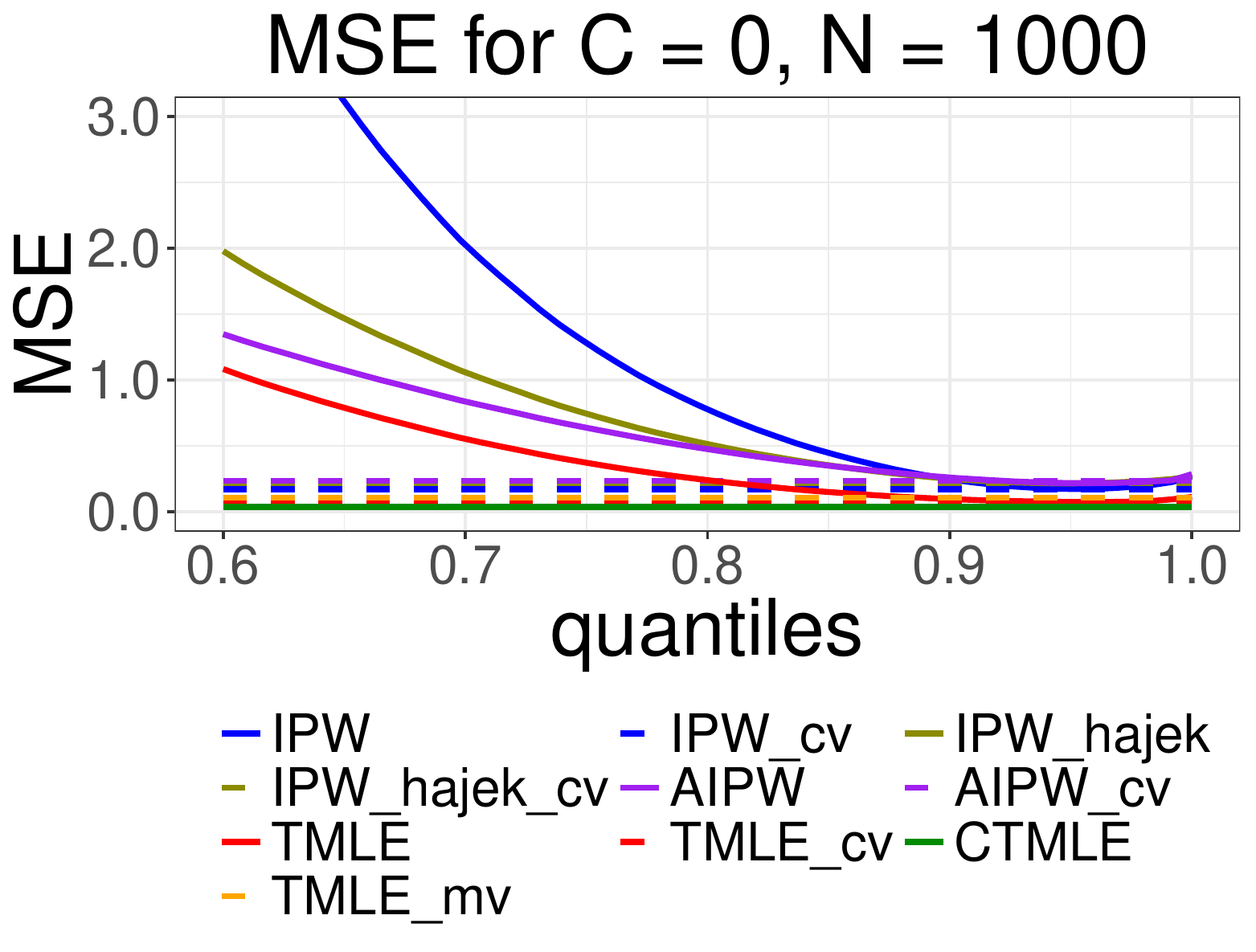}
  \end{subfigure}
  \caption{Comparison of the MSEs for each estimator with $C = 0$.}\label{fig:c1.0}
\end{figure}

First we study the case $C = 0$. When the sample size is 200, small values of $\gamma$ result in high MSE due to bias and large values of $\gamma$ result in high MSE due to variance. 
The optimal cutpoint for different estimator varies. For IPW, IPW-Hajek, and A-IPW, the optimal was about $\gamma=0.9$. It is not surprising to see the vanilla IPW estimator is  the most unstable around $\gamma = 1$. TMLE is the most stable estimator, and it achieved optimal around $\gamma = 0.8$. Among all estimators with cutpoint selected by CV, TMLE performed best, and its MSE was very close to the MSE of C-TMLE. CV-TMLE, MV-TMLE, and C-TMLE have similar performance.

When the sample size is 1000, the optimal cutpoint for all estimators was close to $\gamma =1$. Intuitively, the larger sample size make the variance of estimators smaller, thus less truncation is necessary.  When the sample size is large enough, it would be unnecessary to truncate the PS estimate. All the estimators with cutpoint selected by CV had similar performance. The C-TMLE estimator achieved the best MSE and the CV-TMLE estimator achieved the second best MSE when $N=1000$. For both $N=200$ and $1000$, the C-TMLE estimator was even better than the oracles of all the competing estimators with fixed quantile: the horizontal line for C-TMLE is below all the curves.

\begin{figure}[H]
  \centering
  \begin{subfigure}[b]{0.4\textwidth}
    \includegraphics[width=\textwidth]{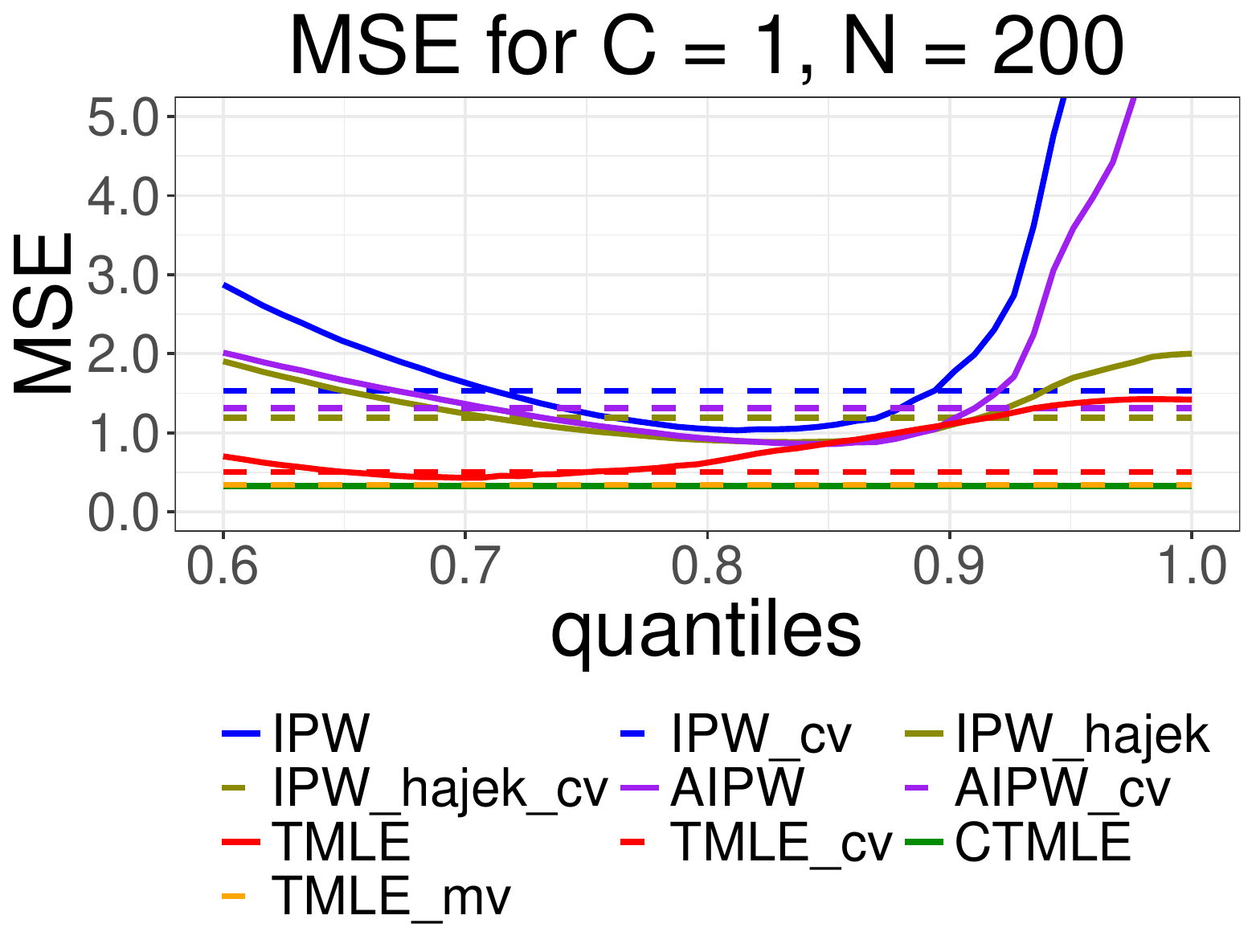}
  \end{subfigure} \hspace{5mm}
  \begin{subfigure}[b]{0.4\textwidth}
    \includegraphics[width=\textwidth]{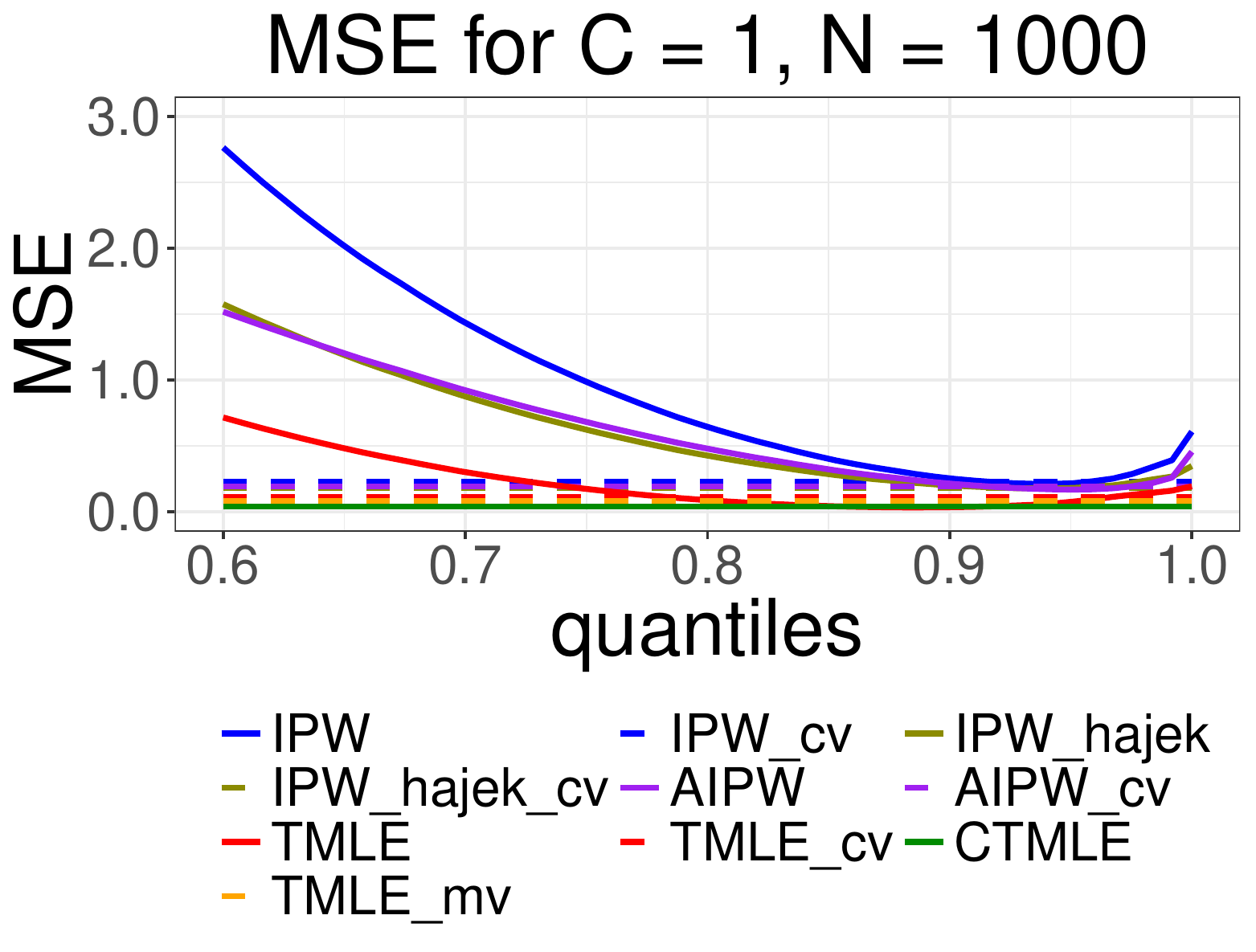}
  \end{subfigure}
  \caption{Comparison of the MSEs for each estimator with $C = 1$.}\label{fig:c1}
\end{figure}

We then set $C=1$ to introduce stronger practical violations of the positivity assumption. For $N=200$, the IPW and A-IPW estimators became more unstable. The corresponding MSEs increased sharply when $\gamma$ increased from 0.85 to 1. This might be because of the unstable inverse weighting in these two estimators. Hajek-type IPW was much more stable for mild truncation, in comparison to IPW and A-IPW. TMLE was more stable and had better  performance compared to the previous estimators. For estimators with adaptive cutpoint selection, TMLE achieved the best performance among all the estimators with cutpoint selected by CV. MV-TMLE and C-TMLE had the best performance among all the estimators.

When $N=1000$, all the estimators have similar performance with the previous case where $C=0, N = 1000$. Due to the relatively large sample size, even the estimators with untruncated PS had satisfactory performance. However, we observe that, different from the case with $N=1000$ and $ C=0$, the  MSE for IPW starts increasing after $\gamma = 0.95$ when $N = 1000, C = 1$, which indicates there are stronger violations of the positivity assumption in this case. In this setting, C-TMLE still achieved the best performance among all estimators and was better than the oracles for all estimators with fixed cutpoint.

\begin{figure}[H]
  \centering
  \begin{subfigure}[b]{0.4\textwidth}
    \includegraphics[width=\textwidth]{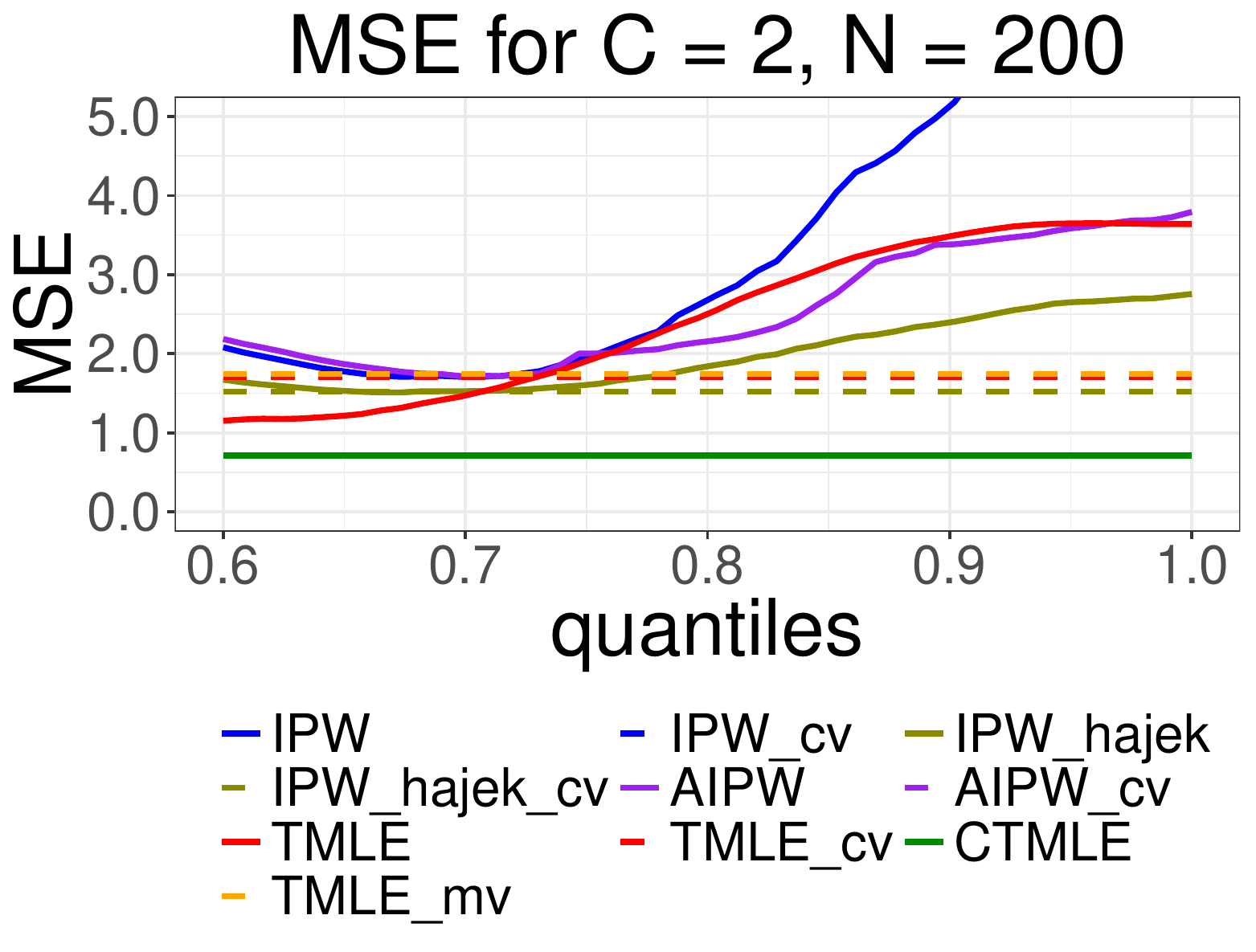}
  \end{subfigure} \hspace{5mm}
  \begin{subfigure}[b]{0.4\textwidth}
    \includegraphics[width=\textwidth]{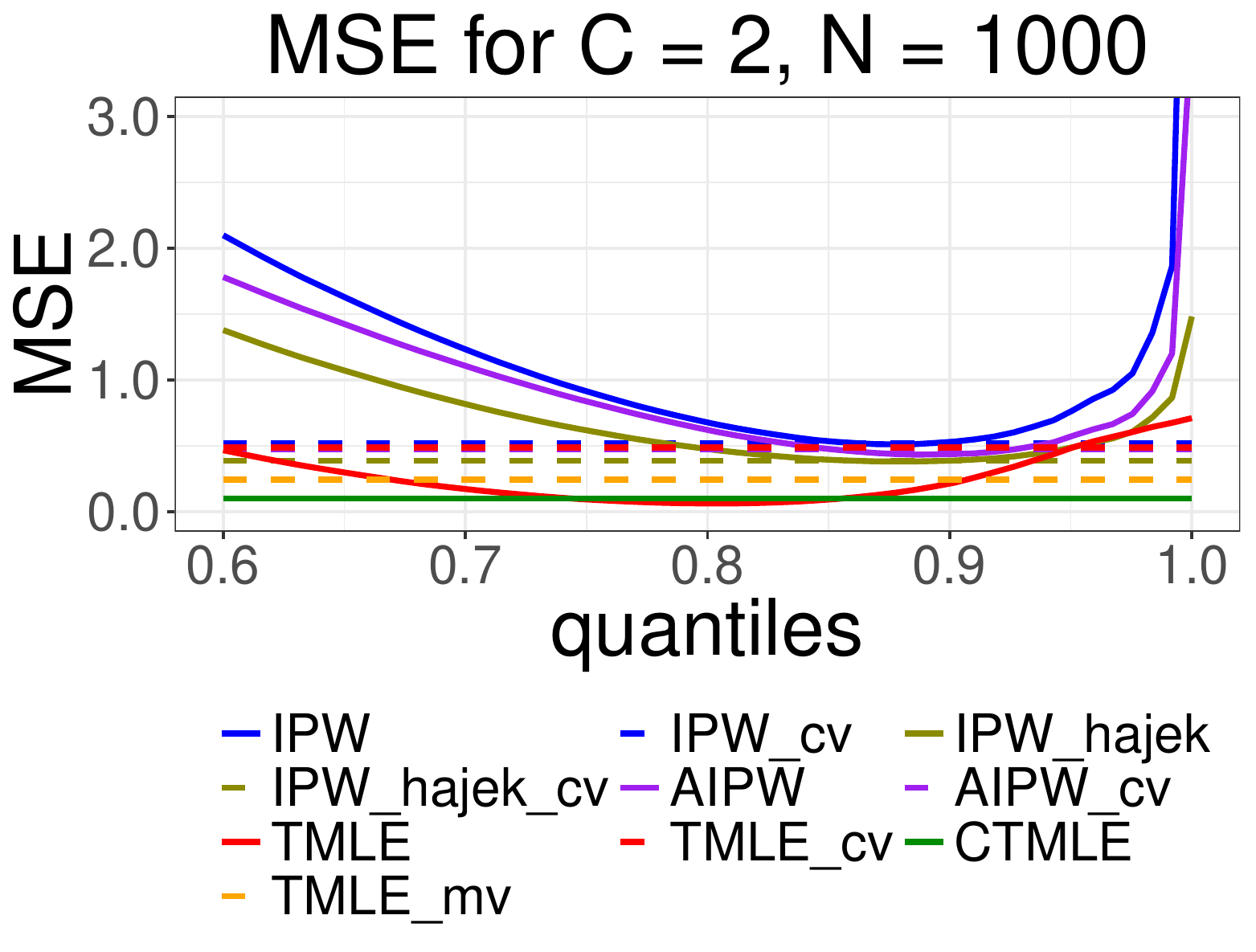}
  \end{subfigure}
  \caption{Comparison of the MSEs for each estimator $C = 2$.}\label{fig:c2}
\end{figure}

Finally we studied the case where the positivity parameter $C= 2$. We could see from figure \ref{fig:g0gnC2} that there was strong practical violation of the positivity assumption, as the distribution of the PS is highly concentrated around 1.
For $N=200$, MSEs for all estimators increased compared to the previous cases where $C = 0,1$. The MSEs for IPW  was out of the bound of the plot when the PS was truncated with large quantile. Hajek-type IPW estimator was much more stable compared to IPW and A-IPW in this case.
TMLE still had satisfactory performance among all the non-adaptive estimators, and the optimal quantile for truncation of TMLE is around $\gamma = 0.6$.
For the estimators with cutpoint selected by CV, Hajek-TMLE achieved the best performance. In this case, where there exist strong practical positivity violations, the gap between C-TMLE estimator and other estimators became larger.

Similar to the previous cases, larger sample size relieved issues from  practical violations of the positivity assumption. When $N = 1000$, the optimal quantile for TMLE truncation increased to around $\gamma = 0.84$, while for all the other non-adaptive estimators the optimal quantile was around $\gamma= 0.9$. The estimators with cutpoint selected by CV had similar performance, with MSE around 0.4, and MV-TMLE estimator had slightly better performance.
C-TMLE estimator had the best performance among all the adaptive estimators.
The oracle for TMLE with fixed cutpoint is slightly better than C-TMLE when $C = 2, N = 1000$, but such optimal cutpoint is unknown in practice.

\subsubsection{The Bias-Variance Trade-off}

We further studied the bias and variance trade-off for each estimator.

\begin{figure}[H]
  \centering
  \begin{subfigure}[b]{0.4\textwidth}
    \includegraphics[width=\textwidth]{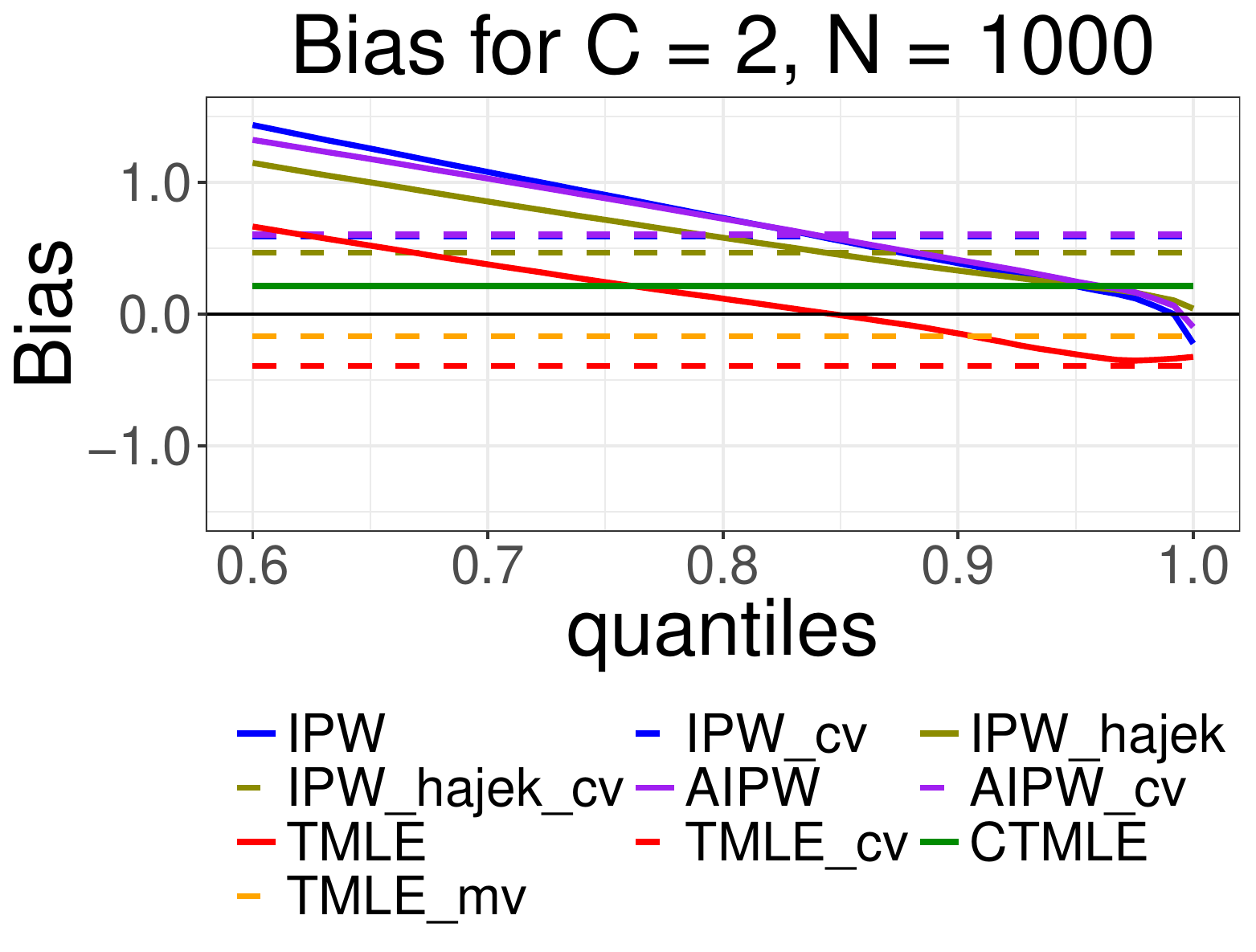}
  \end{subfigure} \hspace{5mm}
  \begin{subfigure}[b]{0.4\textwidth}
    \includegraphics[width=\textwidth]{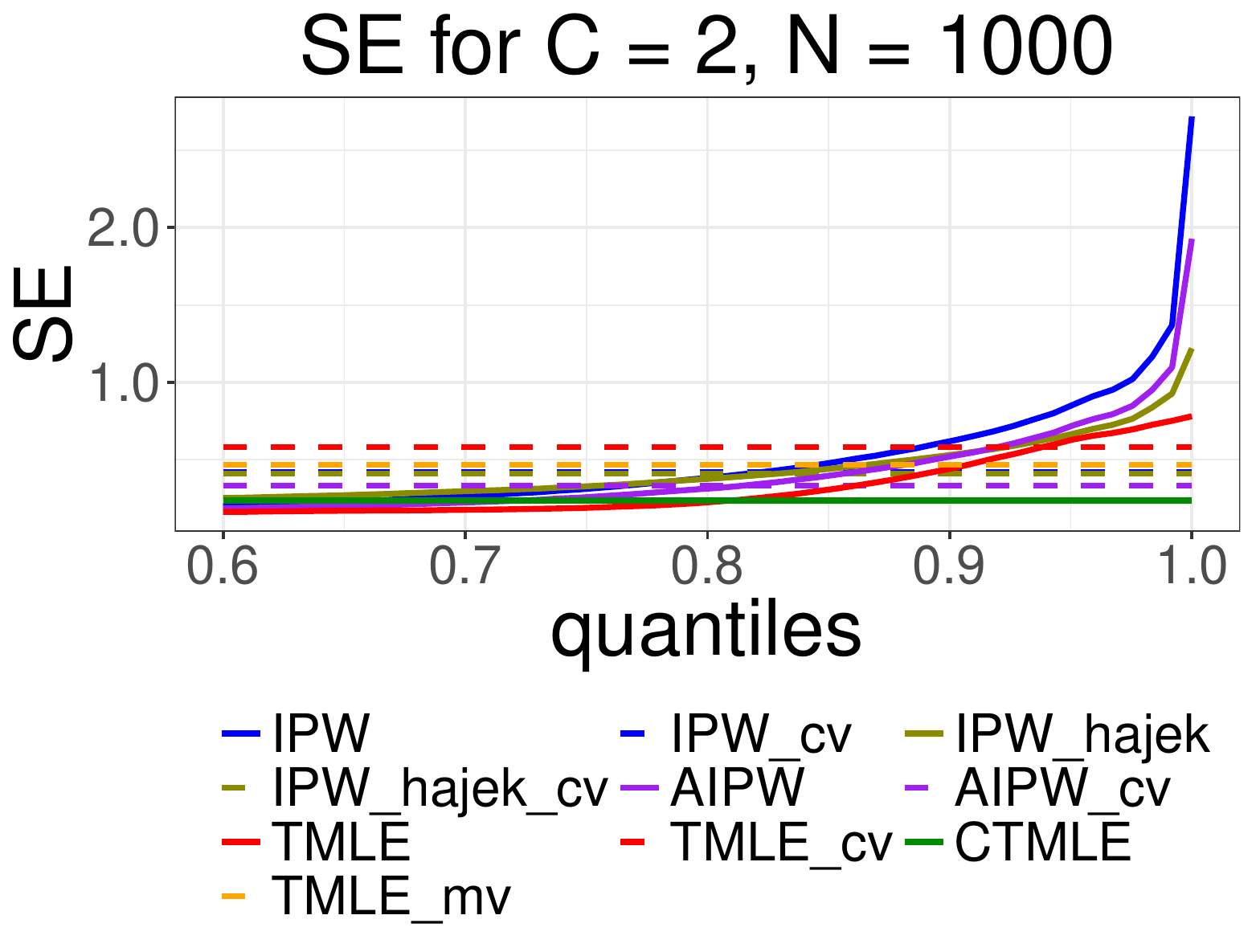}
  \end{subfigure}
  \caption{Comparison of the MSEs for each estimator $C = 2$. For bias plot (left), a horizontal line at 0 is added for better comparison.}\label
          {fig:bv}
\end{figure}

Figure \ref{fig:bv} shows the bias and the standard error (SE) for each estimator. The figure for bias shows that when the cutpoint is increased, IPW, Hajek-type IPW and A-IPW became less biased. The bias of TMLE decreased from positive to 0, and then became negative. This shows  practical violations of positivity would introduce bias for TMLE when no truncation is applied to the PS estimate, even when using the true parametric-model for PS estimation. For the SE, all the estimators with fixed cutpoint shows the same pattern: all the SE increase dramatically with truncation quantile increased from 0.8 to 1.0. For all estimators with adaptive cutpoint selection, C-TMLE achieved both the smallest  SE, and a relatively small bias. In comparison, MV-TMLE and CV-TMLE achieved  small absolute bias, but had overly large variance. Among all estimators using CV for cutpoint selection, TMLE has the best MSE (see figure \ref{fig:c2}).

More details can be found in table~\ref{table:res}.

\begin{table}[ht]
  \centering
  \caption{Detailed results for the data-adaptive truncation methods.}
  \label{table:res}
  \scalebox{0.8}{
\begin{tabular}{r|r||rrr||rrr||rrr|}
  \hline
 N & & \multicolumn{3}{c||}{C = 0} & \multicolumn{3}{c||}{C = 1}   &\multicolumn{3}{c|}{C = 2} \\
  \hline
& Estimator & Bias & SE & MSE & Bias & SE & MSE & Bias & SE & MSE \\ 
  \hline \hline
200& CV-TMLE & 0.243 & 0.575 & 0.389 & 0.325 & 0.619 & 0.488 & -0.067 & 1.471 & 2.163 \\ 
  & MV-TMLE & 0.064 & 0.369 & \textbf{0.140} & 0.117 & 0.409 & \textbf{0.181} & -0.272 & 1.065 & 1.206 \\ 
  & C-TMLE & 0.042 & 0.459 & 0.212 & 0.193 & 0.423 & 0.216 & 0.062 & 0.962 & \textbf{0.927}  \\
\hline \hline
1000 &CV-TMLE & -0.033 & 0.294 & 0.087 & -0.147 & 0.304 & 0.114 & -0.392 & 0.581 & 0.489 \\ 
  &MV-TMLE & 0.106 & 0.310 & 0.107 & 0.017 & 0.291 & 0.084 & -0.167 & 0.467 & 0.245 \\ 
  &C-TMLE & 0.022 & 0.196 & \textbf{0.039} & 0.070 & 0.189 & \textbf{0.040} & 0.214 & 0.237 & \textbf{0.102} \\ 
   \hline
\end{tabular}
}
\end{table}

To further study the estimators with data-adaptive truncation selection, we also compared MSE for each estimator with the positivity parameter $C$ increasing from 0 to 2.

\begin{figure}[H]
  \centering
  \begin{subfigure}[b]{0.4\textwidth}
    \includegraphics[width=\textwidth]{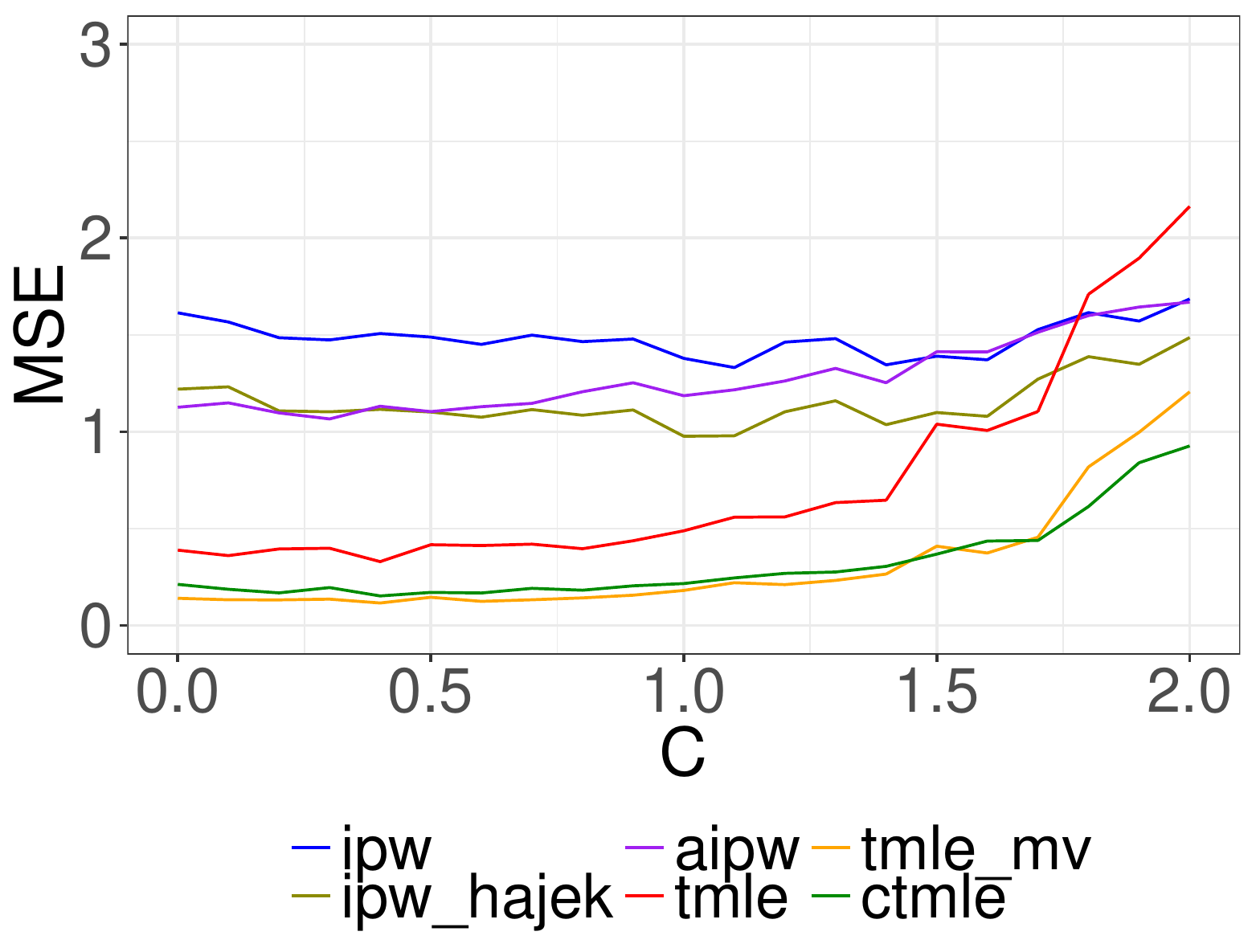}
  \end{subfigure}
  \hspace{5mm}
  \begin{subfigure}[b]{0.4\textwidth}
    \includegraphics[width=\textwidth]{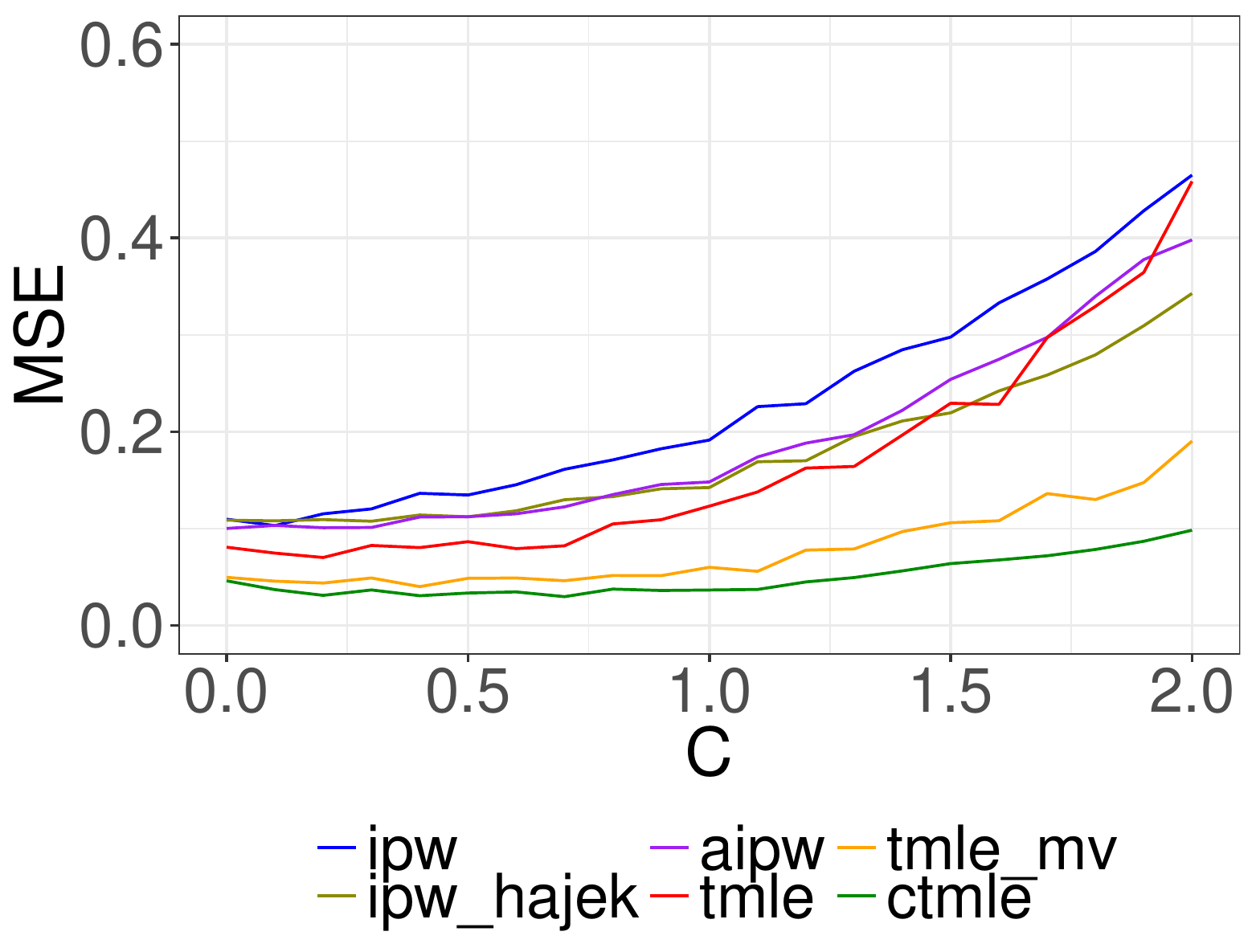}
  \end{subfigure}
  \caption{Comparison of C-TMLE, MV-TMLE, and the other estimators with cutpoint selected by CV. We varied $C$ from 0 to 2. Left: sample size $N =200$. Right: sample size $N = 1000$.}\label{fig:MSEs}
\end{figure}

Figure \ref{fig:MSEs} shows the trend of MSE for each estimator with the positivity parameter $C$ increasing from 0 to 2. C-TMLE kept better performance compared to all the other estimators with cutpoint determined by CV. In addition, the gap between the MSE for C-TMLE and other estimators kept increasing. This suggests that CV is far from the optimal for the cutpoint $\gamma$ selection.

MV-TMLE has good performance when $N = 200$ and $C$ is small. However, in the setting of $N = 200$, when violations of positivity became stronger, its MSE increased dramatically after $C = 1.5$. When the sample size $N = 1000$, it keeps satisfactory performance. However, it is  consistently weaker than C-TMLE across all $C$.

\subsection{Comparison of Cutpoints for CV, MV-TMLE, and C-TMLE}
\label{subsec:comp}

To better understand the difference between the cutpoints $\gamma$ selected by C-TMLE and CV, we study the mean of the quantiles selected for C-TMLE, MV-TMLE and CV. To have a better comparison, we used TMLE estimator with the cutpoint selected by CV  (CV-TMLE) to compare with Positivity-C-TMLE (C-TMLE), and the cutpoint selected by MV (MV-TMLE).

\begin{figure}[H]
  \centering
  \begin{subfigure}[b]{0.3\textwidth}
    \includegraphics[width=\textwidth]{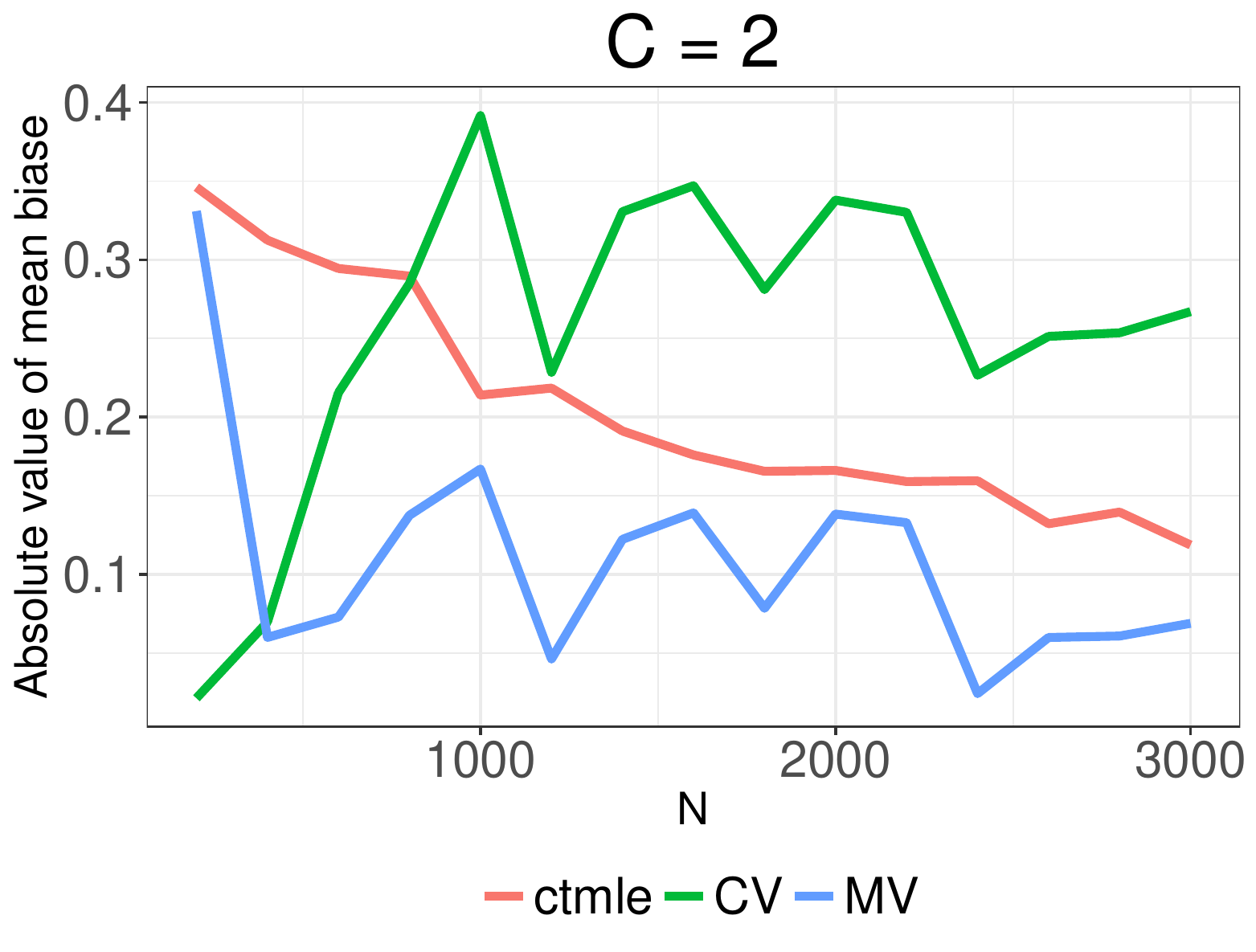}
  \end{subfigure}
  \begin{subfigure}[b]{0.3\textwidth}
    \includegraphics[width=\textwidth]{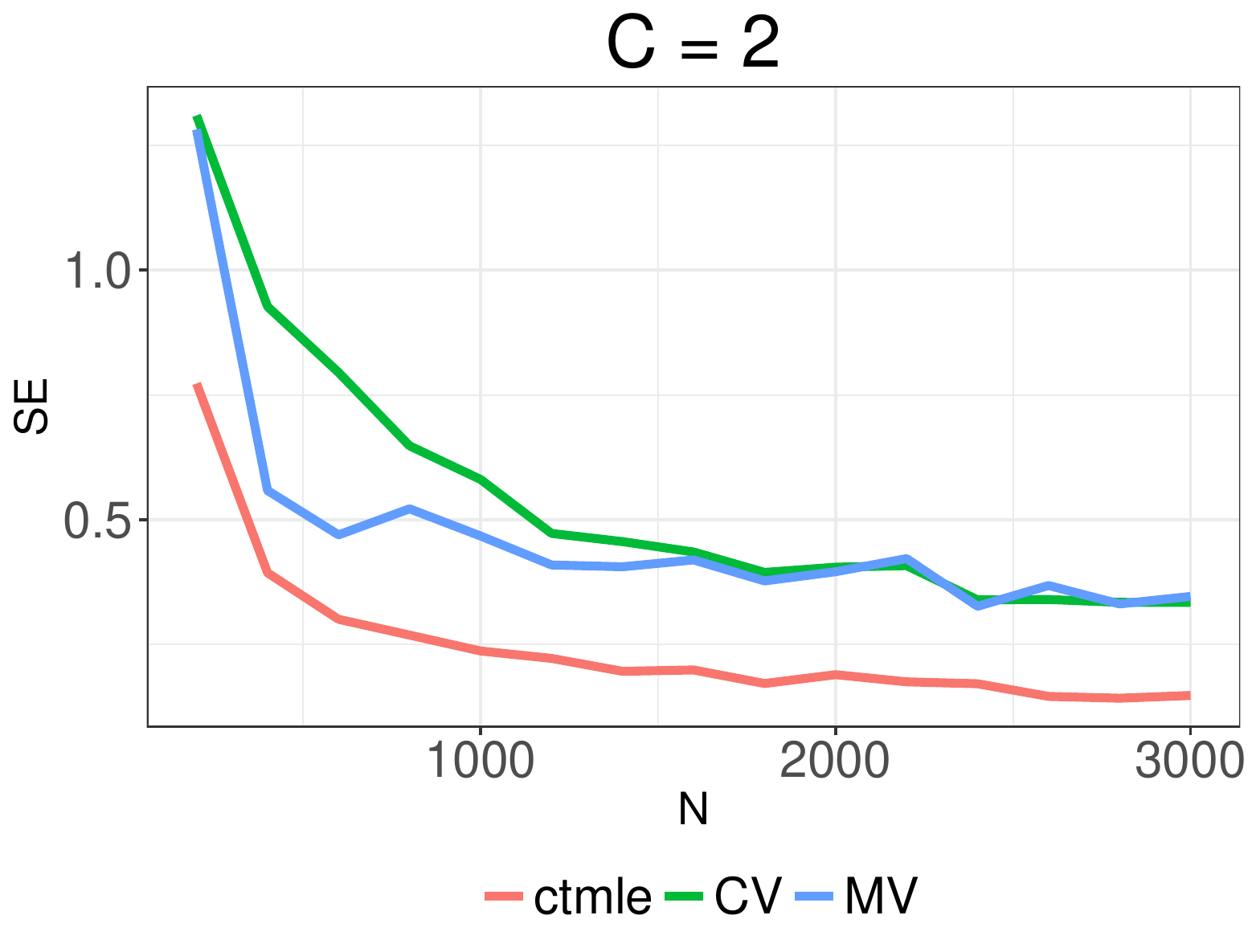}
  \end{subfigure}
  \begin{subfigure}[b]{0.3\textwidth}
    \includegraphics[width=\textwidth]{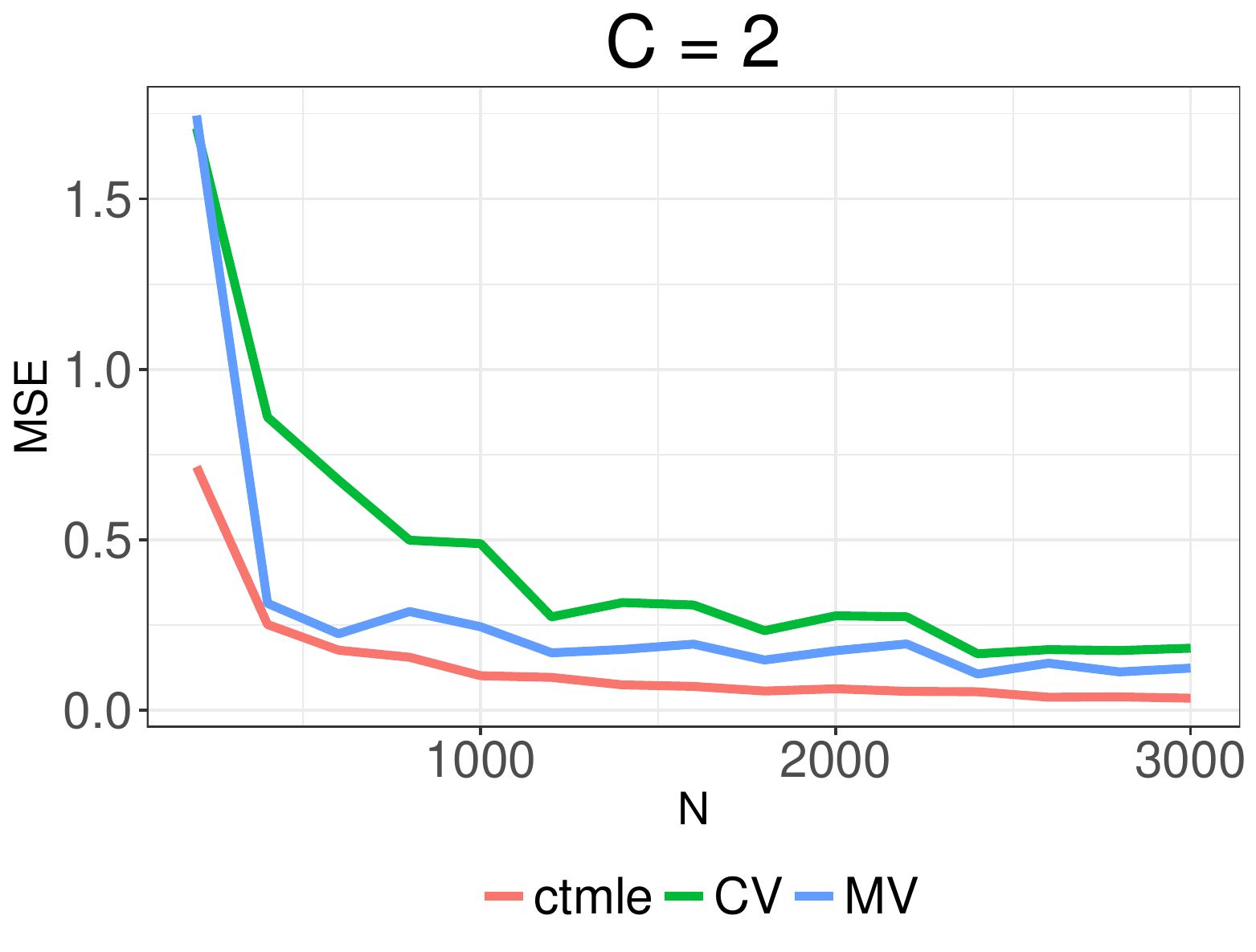}
  \end{subfigure}
  \caption{Fix positivity parameter $C=2$, increase $N = 200$ to $3000$}\label{fig:fixCchangN}
\end{figure}

We first show the absolute mean bias, SE, and MSE for CV-TMLE and C-TMLE with the positivity parameter $C=2$, sample size $N$ changing from 200 to 3000. MSE for both algorithms decreases, which mainly due to the decreasing SE. The absolute mean bias for C-TMLE shows a decreasing trend, but not clear for CV-TMLE. This might because CV is too sensitive to the sample size, and selected too mild truncation (too large cutpoint quantile $\gamma$). In addition, it is interesting to see the bias curves of CV-TMLE and MV-TMLE show very similar pattern.

To better understand why C-TMLE outperforms CV-TMLE, we plot the mean cutpoint selected by CV and C-TMLE.

\begin{figure}[H]
  \centering
  \begin{subfigure}[b]{0.4\textwidth}
    \includegraphics[width=\textwidth]{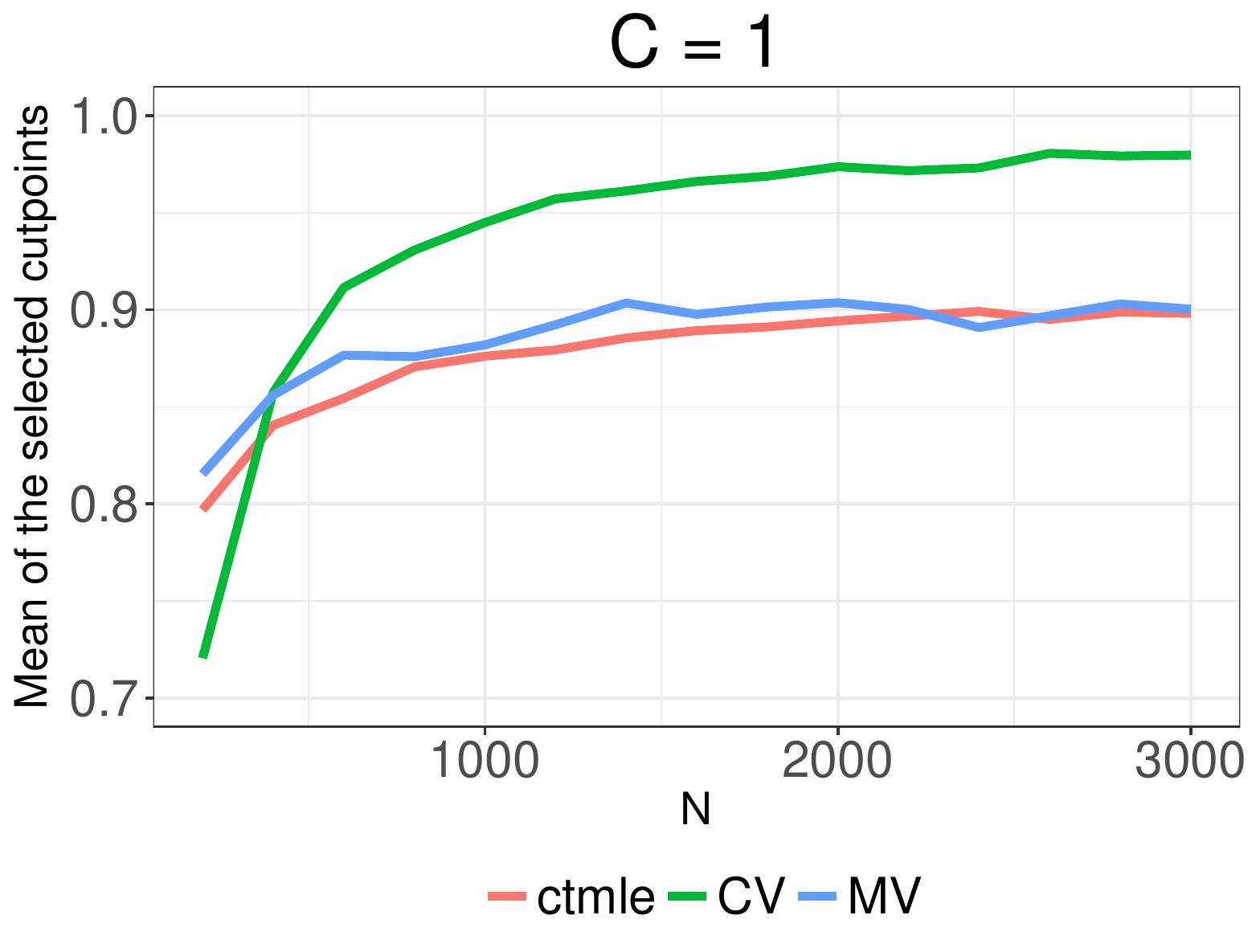}
  \end{subfigure}
  \hspace{5mm}
  \begin{subfigure}[b]{0.4\textwidth}
    \includegraphics[width=\textwidth]{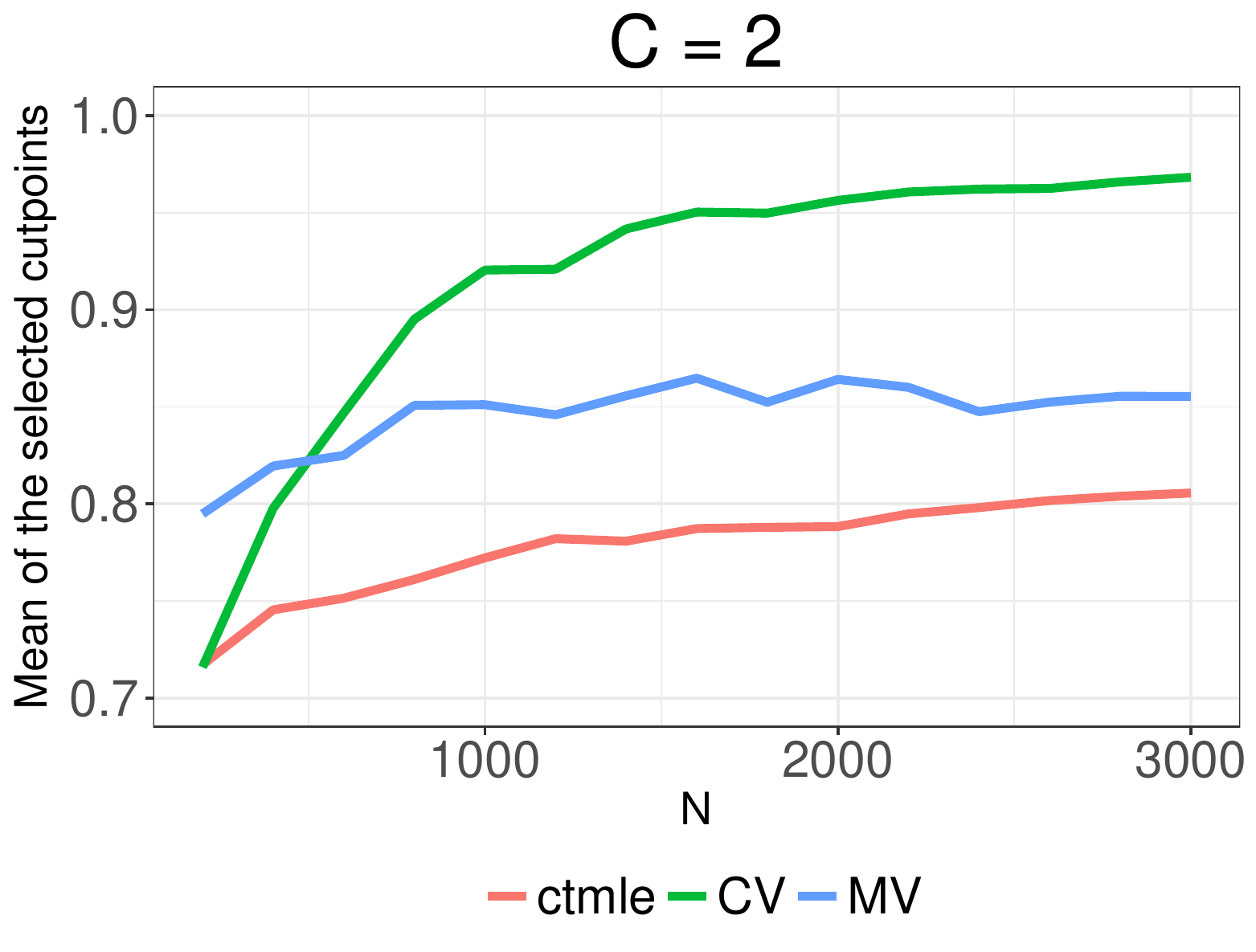}
  \end{subfigure}
  \caption{Mean of selected quantiles by CV, MV-TMLE, and C-TMLE for fixed $C = 1$ (left) and 2 (right), with sample size $N$ increasing from 200 to 3000.}
  \label{fig:quantiles_N}
\end{figure}

Figure \ref{fig:quantiles_N} shows the mean quantile selected by CV and C-TMLE. In this experiment, we fixed $C = 1$ (left) and $C = 2$ (right), with sample size $N$ increasing from 200 to 3000. 
 We observe that CV is more sensitive to $N$ in comparison to MV-TMLE and C-TMLE. The cutpoint increased dramatically from around 0.7 to 0.95, when $N$ increased from 200 to 1000. However, C-TMLE tended to be more conservative. Even when the sample size is very large, it still only truncated at around 90\%. On the other hand, comparing the two figures with $C = 1$ and $ C =  2 $, we could see C-TMLE is much more sensitive to the positivity parameter $C$. In comparison, the lines for CV for $C = 1, 2$ are more similar than the lines for C-TMLE. The cutpoint selected by MV-TMLE is not sensitive either to the sample size, or the positivity parameter.

To better understand their behavior from another perspective, we fixed the sample size $N$ and increased the positivity parameter $C$. 
\begin{figure}[H]
  \centering
  \begin{subfigure}[b]{0.3\textwidth}
    \includegraphics[width=\textwidth]{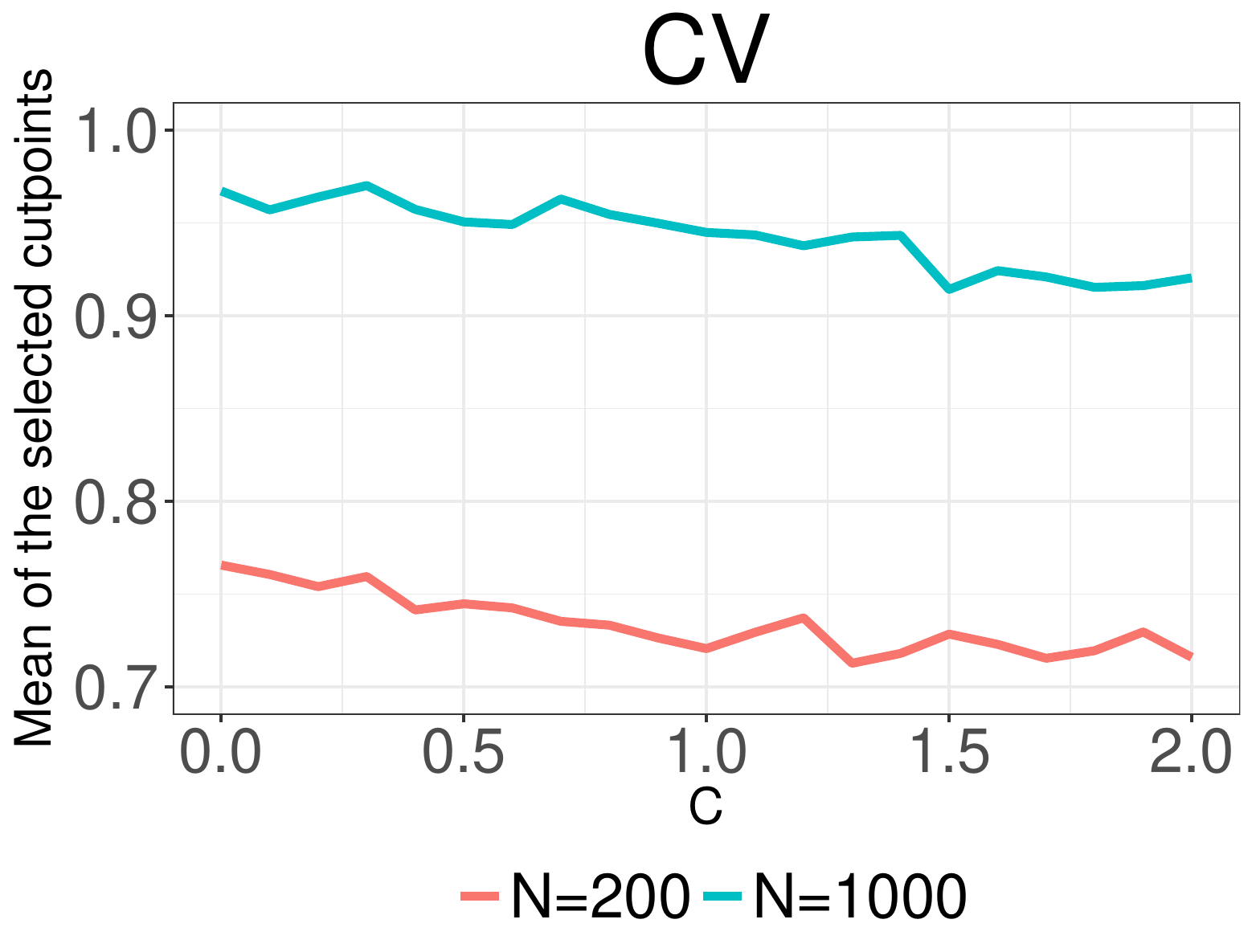}
  \end{subfigure}
  \begin{subfigure}[b]{0.3\textwidth}
    \includegraphics[width=\textwidth]{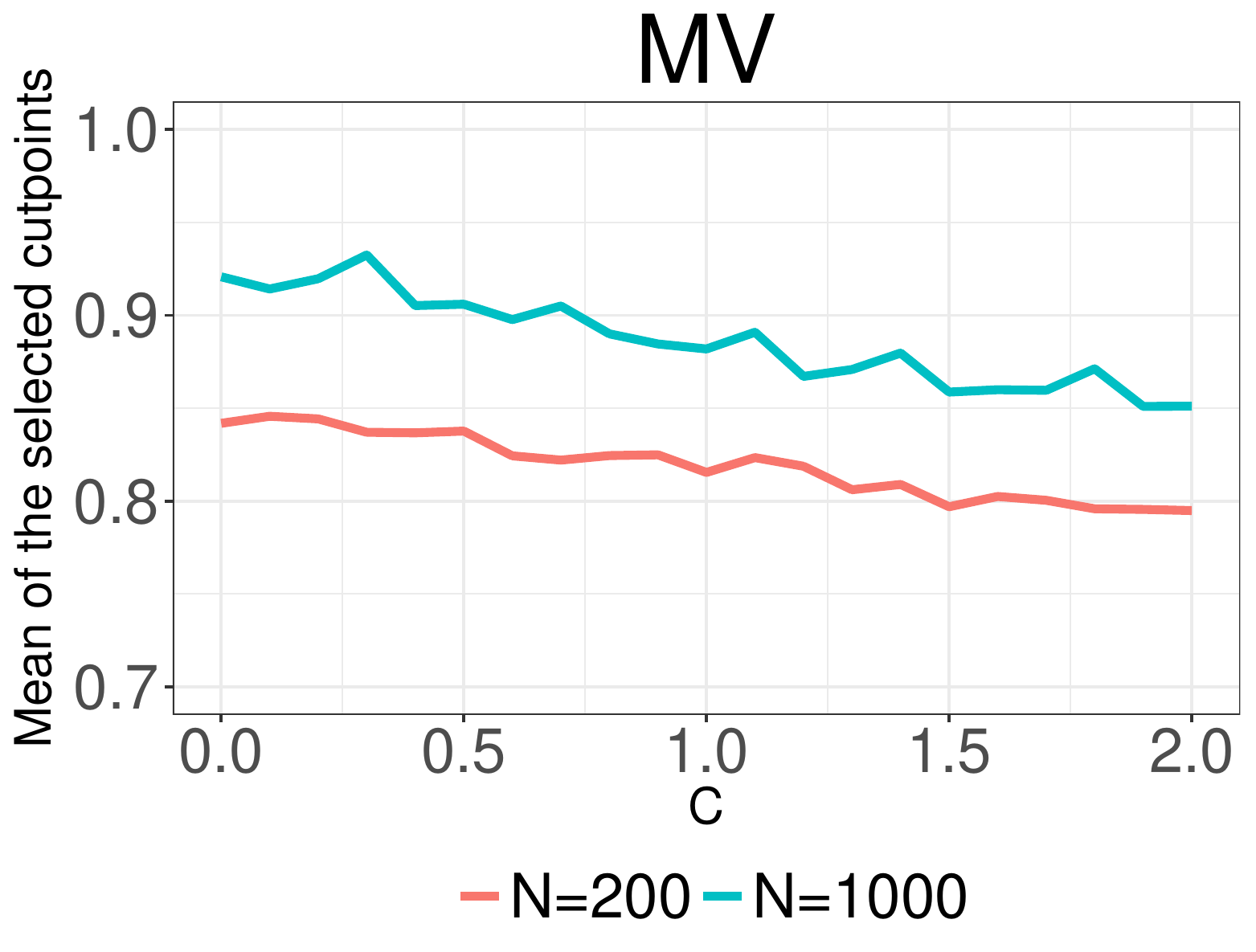}
  \end{subfigure}
  \begin{subfigure}[b]{0.3\textwidth}
    \includegraphics[width=\textwidth]{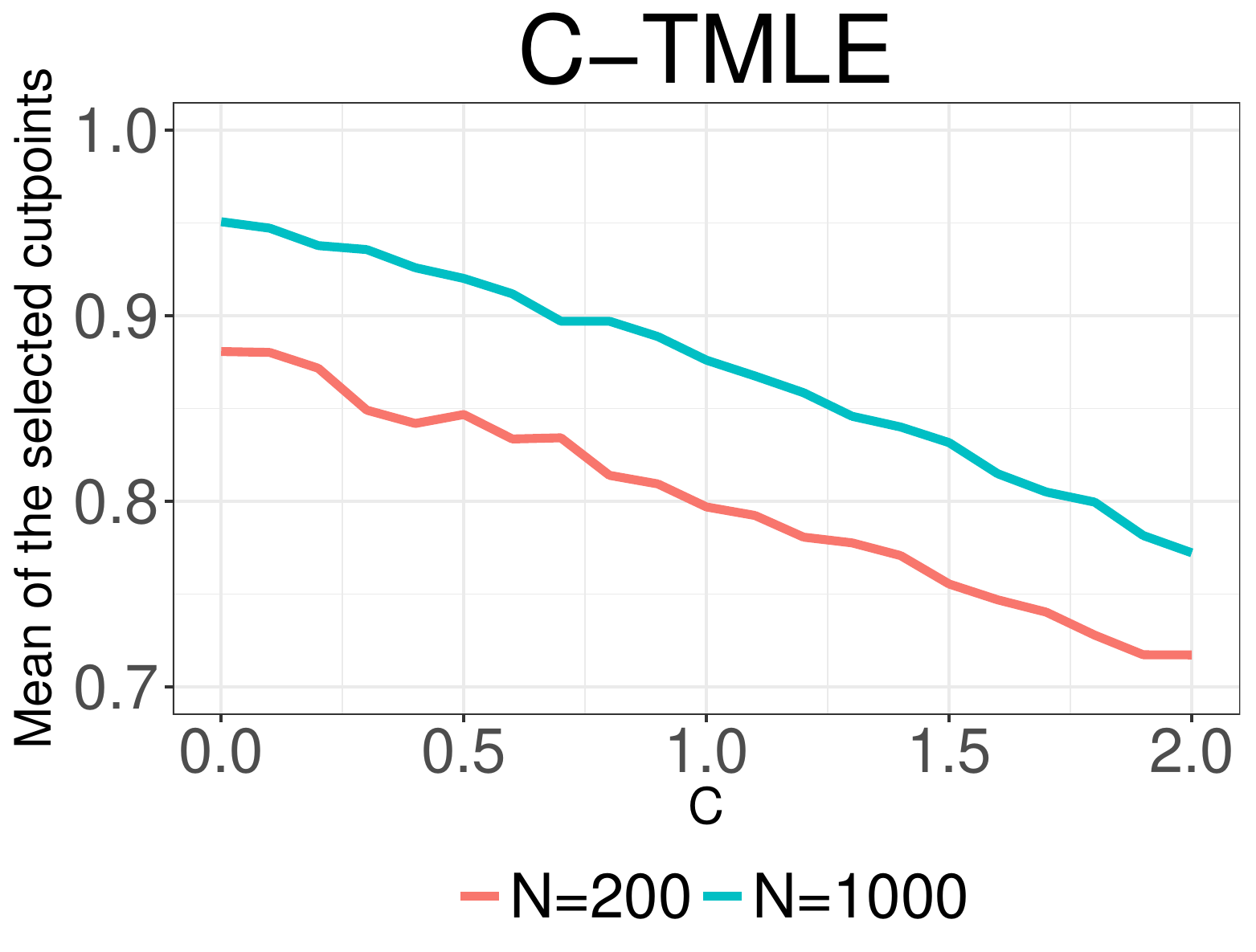}
  \end{subfigure}
  \caption{Mean of selected quantile by CV, MV-TMLE, and C-TMLE, with sample size  $N = 200$ and $n = 1000$ and the positivity parameter $C$ from 0 to 2.}
  \label{fig:quantiles}
\end{figure}

Figure \ref{fig:quantiles} shows the cutpoint selected by C-TMLE is more sensitive to the positivity parameter $C$, as the curves for CV and MV-TMLE are flatter. 
 This  could be explained by the objective function used for CV: the commonly used negative log likelihood loss penalized the observations with:
\begin{equation*}\
  L_2(\bG)(A_i, W_i) = A_i[\log(\bG(W_i))] + (1-A_i) [\log(1-\bG(W_i))]
\end{equation*}
Consider the case where the untreated observations are rare. Then for the untreated observations $A_i=0$, but with high value of the estimated PS, $\bG_n(W_i)$, it would contribute $-\log(1-\bG_n(W_i))$ to the loss function. However, the  performance of the estimators with inverse weighing would suffer more in comparison to the predictive performance of $\bG$, as the inverse of a very small number, $1/\bG(W_i)$, can be much larger/influential. In this sense, the C-TMLE estimator has an attractive property that it determines the cutpoint by minimizing the CV loss for the parameter of interest, instead of the nuisance estimator.

It remains unknown why the cutpoint selected by MV-TMLE is not sensitive to either sample size, or the positivity parameter. Unlike CV, which is model free, MV relies on the choice of the causal estimator. Thus it is possible that the cutpoint would be more sensitive if we switch to a less robust estimator (e.g. IPW estimator).

  For C-TMLE, notice this cutpoint selection is different from the general model selection problem. Unlike the general model selection  (e.g. selection of the regularization parameter $\lambda$ for LASSO), the cutpoint $\gamma$ selection is not closely relevant to the bias-variance trade-off, or smoothness,  of $\bG$, as it only affects the tail distribution of $A \mid W$. The negative log-likelihood would always select little truncation (high cutpoint) as the increasing of bias is faster than the decreasing for variance, as the Kullback-Leibler divergence is not sensitive to predicted probabilities close to 0/1.  Thus it does not fit the general theorem of C-TMLE in \citep{van2017ctmle}, as even $\bG_{n, \gamma}$ selected by CV will yield an asymptotic linear estimator, without suffering from under-smoothing. However, in the finite-sample cases, the Positivity-C-TMLE uses a more targeted criterion in comparison to CV, which leads to a better practical performance.

\subsection{Confidence Intervals} 

In this section, we study the confidence intervals for TMLE and C-TMLE.

\begin{table}[h!]
  \centering
  \caption{Coverage of CI across 200 experiments with sample size 1000 for TMLE and C-TMLE, with the relative mean width of CI to TMLE* in parentheses. Estimators with * use the true SE (provided by a separate Monte Carlo simulation), while the others use the estimated SE. }
  \label{table:ci}
  \scalebox{0.8}{
  \begin{tabular}{|l||l|l|l|l|l|l|} 
    \hline
                             & C = 0 & C =  0.5 &  C = 1 &  C = 1.5 &  C = 2  \\  
    \hline \hline
    CV-TMLE*   & 0.95 (1.00) & 0.93 (1.00) & 0.94 (1.00) & 0.91 (1.00) & 0.91 (1.00) \\
    MV-TMLE*  & 0.95 (1.05) & 0.94 (0.92) & 0.96 (0.95) & 0.94 (0.77) & 0.94 (0.80)\\
    C-TMLE*    & 0.95 (0.68) & 0.94 (0.70) & 0.94 (0.69) & 0.92 (0.51) & 0.92 (0.46)\\
    \hline \hline
    CV-TMLE    & 0.85 (0.82) & 0.85 (0.79) & 0.76 (0.69) & 0.67 (0.57) & 0.61 (0.51)\\
    MV-TMLE   & 0.92 (0.83) & 0.88 (0.71) & 0.80 (0.62) & 0.79 (0.49) & 0.67 (0.40)\\
    C-TMLE     & 0.95 (0.60) & 0.88 (0.60) & 0.84 (0.49) & 0.82 (0.39) & 0.70 (0.34)\\
    \hline
  \end{tabular}
  }
\end{table}

Table \ref{table:ci} shows the average coverage and length of confidence intervals.  The positivity would also influence the estimation of the variance of the estimators. To better understand behavior of the two estimators, we studied two settings. In the first setting, we used the true SE,  $\SE(\Psi_n)$, of the CV-TMLE, MV-TMLE, and C-TMLE (computed by a Monte Carlo simulation), and applied it to construct the CIs: $[\Psi_n - 1.96 \cdot \SE(\Psi_n), \Psi_n + 1.96  \cdot\SE(\Psi_n)]$. In the second case, we applied the  estimated SE, $\hat{\SE}(\Psi_n)$, to construct CIs  $[\Psi_n - 1.96  \cdot\hat{\SE}(\Psi_n), \Psi_n + 1.96  \cdot\hat{\SE}(\Psi_n)]$ for both CV-TMLE, MV-TMLE, and C-TMLE.

First, we observe the TMLE* had much larger variance but smaller bias compared to C-TMLE in this experiment ($C = 2, N = 1000$). The large variance of TMLE helps the coverage for its CI, if we know the true variance (which is not possible). C-TMLE selects the cutpoint by optimizing the bias-variance trade-off to the MSE of the targeted parameter, and thus  introduces more bias to reduce the variance in order to achieve better MSE. This is also shown in figure \ref{fig:quantiles}, where in sample size 1000, CV would on average truncate with a larger quantile. The overly large variance causes a much wider CI, which leads to the satisfactory coverage for TMLE*, though this makes the TMLE estimator less efficient.

However, as the true variance of the estimator is unknown in practice, CIs usually rely on the estimation of the variance. We observe that the variance of  CV-TMLE, MV-TMLE,  and C-TMLE estimator was underestimated in our experiments. For all the estimators, the estimated variance were smaller than the true variances. Extreme weights in the clever covariates $H(A,W)$ cause large variance of the influence curve, thus makes it challenge to estimate the variance of the estimator. The variance estimator for the Positivity-C-TMLE estimator is less biased than the variance estimators for the  CV-TMLE and MV-TMLE estimator. The ratio of the mean estimated SE to the corresponding true SE is about $0.88, 0.86, 0.71, 0.76, 0.74$ for $C = 0, 0.5, 1, 1.5, 2$, respectively. While for CV-TMLE and MV-TMLE, the ratio is much smaller. The ratio of the mean estimated SE to the corresponding true SE for the CV-TMLE estimator is about  $0.82, 0.79, 0.69, 0.57, 0.51$,  and for the MV-TMLE estimator is about $0.79, 0.77, 0.65, 0.64, 0.49$, for $C = 0, 0.5, 1, 1.5, 2$ respectively. This explains why CV-TMLE and MV-TMLE  had worse CI coverage than C-TMLE.

\section{Conclusion}

In this study, we proposed the Positivity-C-TMLE algorithm for  adaptive truncation of the PS  to address the issues from practical violations of the positivity assumption. We also designed the simulations to evaluate and to help understand this novel estimator. We have the following conclusions:

\begin{itemize}

\item It is reasonable to believe that the optimal cutpoint varies significantly for different estimators. The Positivity-C-TMLE algorithm was designed for selecting the optimal cutpoint for TMLE, which might be the key point for its outstanding performance in the simulation.

\item As discussed in subsection \ref{subsec:comp}, the negative log-likelihood  function  $L_2$ for $\bG$ is not a good objective function for selecting $\gamma$. Positivity-C-TMLE selects $\gamma$ directly based on the targeted parameter, which is another important factor in its success in the simulation.

\item The cutpoint selected by Positivity-C-TMLE is more sensitive to the positivity parameter $C$ than the cutpoint selected by CV. The cutpoint selected by CV is sensitive to the sample size $N$, but not for the positivity parameter $C$. The cutpoint selected by MV-TMLE is not sensitive either to $N$, or to $C$.

\item For Positivity-C-TMLE, the variance is under-estimated in the simulation, especially when practical violation of the positivity assumption is strong. Though the variance estimator for Positivity-C-TMLE is less biased than the one for CV-TMLE or MV-TMLE,  a more conservative variance estimator is necessary to build a more reliable confidence interval for finite-sample study. One potential solution is to use the robust variance estimator \citep{schwab2014ltmle}. We leave this for future work.

  \item MV-TMLE has similar performance to  C-TMLE when the sample size is large, or when practical violations are mild. However, in small samples with strong positivity violations (e.g. $N = 200, C = 2$), C-TMLE has much better performance than MV-TMLE.

\end{itemize}


There are several potential future extensions of this study. First, we only studied the case where the propensity score is estimated by a correctly specified parametric model. In other words, the failure of the estimators in the simulations are only from the practical violations of the positivity assumption, rather than  model misspecification. It is important to investigate the behavior of each adaptive truncation method when the estimator for $\bG_0$ is misspecified.
In addition, this C-TMLE procedure could be extended to other data structure, like longitudinal data. We leave this for future work.


\bibliographystyle{abbrvnat}
\bibliography{references}

\end{document}